

\input epsf.tex
\def\thetitle{Electroweak Effective Lagrangians}

\def\title{Review paper on effective lagrangians}

\def\theabstract{{
In this paper I review several aspects of the
use of effective lagrangians in (mainly) electroweak physics.
The conditions under which this approach is reliable and useful,
as well as the limitations of the formalism are detailed. Various
applications are also presented.}}

%
%
%
\def\and{{\it\&}}
\def\half{{1\over2}}

\def\quarter{{1\over4}}

\def\gesim{\,{\raise-3pt\hbox{$\sim$}}\!\!\!\!\!{\raise2pt\hbox{$>$}}\,}
\def\lesim{\,{\raise-3pt\hbox{$\sim$}}\!\!\!\!\!{\raise2pt\hbox{$<$}}\,}
\def\boldoverdot{\,{\raise6pt\hbox{\bf.}\!\!\!\!\>}}

\def\ie{{\it i.e.}}

\def\ibid{{\it ibid.}}
\def\etal{{\it et. al.}}
\def\acal{{\cal A}}

\def\dcal{{\cal D}}

\def\gcal{{\cal G}}

\def\ical{{\cal I}}

\def\lcal{{\cal L}}

\def\ocal{{\cal O}}

\def\wcal{{\cal W}}

\def\WW{{\bf W}}

\def\pibf{{\pmb{$\pi$}}}

\def\sigbf{{\pmb{$\sigma$}}}

\def\vev{vacuum expectation value}
\def\irrep{irreducible representation}

\def\tr{ \hbox{tr}}

\def\diag{\hbox{\diag}}
\def\sm{Standard Model}

\def\gev{~\hbox{GeV}}
\def\tev{~\hbox{TeV}}

%
%
%
\def\leaderfill{\leaders\hbox to 1 em{\hss.\hss}\hfill}
\def\inbox#1{\vbox{\hrule\hbox{\vrule\kern5pt
     \vbox{\kern5pt#1\kern5pt}\kern5pt\vrule}\hrule}}
\def\sqr#1#2{{\vcenter{\hrule height.#2pt
      \hbox{\vrule width.#2pt height#1pt \kern#1pt
         \vrule width.#2pt}
      \hrule height.#2pt}}}
\def\square{\mathchoice\sqr56\sqr56\sqr{2.1}3\sqr{1.5}3}
\def\today{\ifcase\month\or
  January\or February\or March\or April\or May\or June\or
  July\or August\or September\or October\or November\or December\fi
  \space\number\day, \number\year}
\def\pmb#1{\setbox0=\hbox{#1}%
  \kern-.025em\copy0\kern-\wd0
  \kern.05em\copy0\kern-\wd0
  \kern-.025em\raise.0433em\box0 }
\def\up#1{^{\left( #1 \right) }}
\def\lowti#1{_{{\rm #1 }}}
\def\inv#1{{1\over#1}}
\def\su#1{{SU(#1)}}
\def\ui{U(1)}
\def\antes{}
\def\despues{.}
%

%
\def\sumprime_#1{\setbox0=\hbox{$\scriptstyle{#1}$}
  \setbox2=\hbox{$\displaystyle{\sum}$}
  \setbox4=\hbox{${}'\mathsurround=0pt$}
  \dimen0=.5\wd0 \advance\dimen0 by-.5\wd2
  \ifdim\dimen0>0pt
  \ifdim\dimen0>\wd4 \kern\wd4 \else\kern\dimen0\fi\fi
\mathop{{\sum}'}_{\kern-\wd4 #1}}
\def\sumbiprime_#1{\setbox0=\hbox{$\scriptstyle{#1}$}
  \setbox2=\hbox{$\displaystyle{\sum}$}
  \setbox4=\hbox{${}'\mathsurround=0pt$}
  \dimen0=.5\wd0 \advance\dimen0 by-.5\wd2
  \ifdim\dimen0>0pt
  \ifdim\dimen0>\wd4 \kern\wd4 \else\kern\dimen0\fi\fi
\mathop{{\sum}''}_{\kern-\wd4 #1}}
\def\sumtriprime_#1{\setbox0=\hbox{$\scriptstyle{#1}$}
  \setbox2=\hbox{$\displaystyle{\sum}$}
  \setbox4=\hbox{${}'\mathsurround=0pt$}
  \dimen0=.5\wd0 \advance\dimen0 by-.5\wd2
  \ifdim\dimen0>0pt
  \ifdim\dimen0>\wd4 \kern\wd4 \else\kern\dimen0\fi\fi
\mathop{{\sum}'''}_{\kern-\wd4 #1}}
%
%
\newcount\chapnum

\def\chap#1{\clearsect\clearprob
\global\advance\chapnum by 1 \par\vskip .5 in\par%
\centerline{{\bigboldiii\antes\the\chapnum\despues\ #1}}}
\newcount\sectnum
\def\clearsect{\sectnum=0}
\def\sect#1{\clearprob\global\advance\sectnum by 1 \par\vskip .25 in\par%
\noindent{\bigboldii\the\chapnum.\the\sectnum:\ #1}\nobreak}
\newcount\yesnonum

\def\verify{\global\advance\yesnonum by 1{\bigboldi (VERIFY!!)}}
\def\tocheck{\par\vskip 1 in{\bigboldv TO VERIFY: \the\yesnonum\ ITEMS.}}
\newcount\notenum

\def\note#1{\global\advance\notenum by 1{ \bf $<<$ #1 $>>$ } }
\def\noteout{\par\vskip 1 in{\bigboldiv NOTES: \the\notenum.}}
\newcount\borrownum

\def\borrow{\global\advance\borrownum by 1{\bigboldi BORROWED BY:\ }}
\def\borrowed{\par\vskip 0.5 in{\bigboldii BOOKS OUT:\ \the\borrownum.}}
\newcount\refnum

\def\ref#1{\global\advance\refnum by 1\item{\the\refnum.\ }#1}
\def\stariref#1{\global\advance\refnum by 1\item{%
               {\bigboldiv *}\the\refnum.\ }#1}
\def\stariiref#1{\global\advance\refnum by 1\item{%
               {\bigboldiv **}\the\refnum.\ }#1}
\def\stariiiref#1{\global\advance\refnum by 1\item{
               {\bigboldiv ***}\the\refnum.\ }#1}
\newcount\probnum
\def\clearprob{\probnum=0}
\def\prob{\global\advance\probnum by 1 {\medskip $\triangleright$\
\undertext{{\sl Problem}}\ \the\chapnum.\the\sectnum.\the\probnum.\ }}
\newcount\probchapnum

\def\probchap{\global\advance\probchapnum by 1 {\medskip $\triangleright$\
\undertext{{\sl Problem}}\ \the\chapnum.\the\probchapnum.\ }}
\def\undertext#1{$\underline{\smash{\hbox{#1}}}$}
%
%
%

\def\UCR{
{{\it University of California at Riverside\break
                  Department of Physics\break
                  Riverside, California 92521--0413; U.S.A. \break
                  \bit}}}

\def\bit{{E-Mail address: wudka{\rm@}phyun0.ucr.edu}}
%
%
%
%
\font\sanser=cmssq8

%

%

%
\font\bigboldi=cmbx10 scaled\magstep1
\font\bigboldii=cmbx10 scaled\magstep2
\font\bigboldiii=cmbx10 scaled\magstep3
\font\bigboldiv=cmbx10 scaled\magstep4
\font\bigboldv=cmbx10 scaled\magstep5
\font\eightrm=cmr8

\input phyzzx
\Tenpoint
\PHYSREV
\def\square{\mathchoice\sqr56\sqr56\sqr{2.1}3\sqr{1.5}3}
\def\vev{vacuum expectation value}
\rightline{UCRHEP-T121}
{\titlepage
\vskip -.2 in
\title{ {\bigboldiii \thetitle}}
\doublespace
\author{{ Jos\'e Wudka }}
\address{\UCR}
\abstract
\bigskip
\singlespace
\theabstract
\endpage}


\def\bfit#1{\pmb{\it#1}}
\def\cw{ c \lowti{ W }}
\def\ctw{ c \lowti{2 W }}
\def\dec{decoupling scenario}
\def\ibos{{I_\phi}}
\def\ifer{{I_\psi}}
\def\ila{{\inv{\Lambda^2}}}
\def\lat{ \Lambda \lowti{TeV}}
\def\hang{\hangindent\parindent}
\def\iitem{\par\hang\textindent}
\def\leff{\lcal \lowti{eff}}
\def\liv{ \lcal \up 4 }
\def\lb{\Lambda_\phi}
\def\lf{\Lambda_\psi}
\def\mh{ m \lowti{ H }}
\def\mw{ m\lowti{W}}
\def\mz{ m \lowti{ Z }}
\def\ndec{non-decoupling scenario}
\def\npb{{\it Nucl. Phys.} {\bf B}}
\def\plb{{\it Phys. Lett.} {\bf B}}
\def\pr{{\it Phys. Rev.}}
\def\prl{{\it Phys. Rev. Lett.}}
\def\prd{{\it Phys. Rev.} {\bf D}}
\def\sw{ s \lowti{W}}
\def\sqr#1#2{{\vcenter{\hrule height.#2pt
      \hbox{\vrule width.#2pt height#1pt \kern#1pt
         \vrule width.#2pt}
      \hrule height.#2pt}}}
\def\square{\mathchoice\sqr56\sqr56\sqr{2.1}3\sqr{1.5}3}
\def\thew{ \theta \lowti{ W }}
\def\tw{ t \lowti{ W } }
\def\vev{vacuum expectation value}
\def\zphys{{\it Z. Phys.} {\bf C}}

\REF\polchinsky{J. Polchinski, lectures presented at {\it TASI 92},
                Boulder, CO, Jun 3-28, 1992.}
\REF\wilson{ K.G. Wilson and J. Kogut, {\it Phys. Rep.} 12 (1974) 75.}
\REF\weinberg{S. Weinberg, {\it Physica} 96{\bf A} (1979) 327.}
\REF\georgi{H. Georgi, \npb361 (1991) 339; \ibid 363 (1991) 301.}
\REF\martyjose{M.B. Einhorn and J. Wudka, in {\it Workshop on Electroweak
               Symmetry Breaking}, Hiroshima, Nov. 12-15 (1991);
               in {\it Yale Workshop on Future Colliders,} Oct. 2-3 (1992).
               M.B. Einhorn, in {\it Conference on Unified Symmetry
               in the Small and in the Large,} Coral Gables, Fl, Jan.
               25-27 (1993).
               M.B. Einhorn, plenary talk given at the {\it Workshop on
               Physics and Experimentation with Linear $e^+e^-$ Colliders},
               Waikoloa, Hawaii, April 26-30, 1993. Univ. of Michigan
               report UM-TH-93-17, to be published in the proceedings.
               J. Wudka, in {\it Electroweak Interactions and Unified
               Theories,} XXVIII Recontres de Moriond Les Arcs, Savoie,
               France, March 13-20 (1993).}
\REF\derujula{A. De R\'ujula \etal, \npb384 (1992) 3.}
\REF\londonburgessi{C.P. Burgess and D. London \prl\ {\bf69} (1992) 3428.}
\REF\casalbuoni{R. Casalbuoni \etal, \plb155 (1985) 95; \npb282 (1987) 235.}
\REF\gl{J. Gasser and H. Leutwyler, \npb250 (1985) 465;
        {\it Ann. Phys.} {\bf158} (1984) 142.}
\REF\pich{A. Pich, lectures presented at the {\it V Mexican School of
          Particles and Fields}, Guanajuato, M\'exico, Dec. 1992.}
\REF\langacker{P. Langacker \etal, {\it Rev. Mod. Phys.}64 (1992) 87.}
\REF\herrero{A. Dobado \etal, \plb235 (1990) 129; \zphys50 (1991) 465.
             M.J. Herrero, in  {\it 1st Int. Triangle  Workshop:
             Standard Model and Beyond: From LEP to UNK and LHC},
             Dubna, USSR, Oct 1-5, 1990; and references therein.}
\REF\bagger{J. Bagger \etal, Fermilab report
             FERMILAB-PUB-93-040-T (unpublished).
             (Bulletin Board: hep-ph@xxx.lanl.gov - 9306256)}
\REF\baggerdawson{J. Bagger \etal, \npb339 (1993) 364.}
\REF\chanowitzandgaillard{J.M. Cornwall \etal, \prd10 (1974) 1145, \ibid {\bf
D}11 (1975) 972.
          C.E. Vayonakis, {\it Lett. Nuov. Cim.} 17 (1976) 383.
          M. Chanowitz and M.K. Gaillard, \npb261 (1985) 379.
          B.W. Lee \etal, \prd16 (1985) 1519.}
\REF\bargerandphillips{J. Barger and R.J.N. Phillips, lectures
                       presented by J. Barger at the {\it VII Jorge
                       Andr\'e Swieca Summer School: Particles and
                       Fields}, Sao Paulo, Brazil, 10-23 Jan 1993.}
\REF\bauzeppenfeld{U. Baur and D. Zeppenfeld, \plb201 (1988) 383.
                   U. Baur and E.L. Berger, \prd47 (1993) 4889.}%
\REF\hagiwara{ K. Hagiwara \etal, \npb282 (1987) 253.}
\REF\goldenpriv{M. Golden, private communication.}
\REF\burgessandlondon{C.P. Burgess and D. London, McGill University report
                      MCGILL-92-04 (unpublished);
                      (Bulletin Board: hep-ph@xxx.lanl.gov - 9203215).}
\REF\gates{S.J. Gates, Jr. \etal {\it Superspace} (Benjamin/Cummings,
           Reading, MA, 1983).}
\REF\delbourgo{R. Delbourgo and G. Thompson \prl57:21, 2601 (1986).}
\REF\grossekneterandkogerler{C. Grosse-Knetter and R. K\"ogerler,
                             \prd48 (1993) 2865.}
\REF\frere{J.-M. Fr\`ere \etal, \plb292 (1992) 348.}
\REF\thooft{G. t'Hooft, in {\it Recent Developments
            in Gauge Theories,} G. t'Hooft \etal eds.
            (Plenum Press, New York 1980).}
\REF\decoupling{T. Appelquist and J. Carazzone, \prd11 (1975) 2856.
                J.C. \etal, \prd18 (1978) 242.
                For a pedagogical introduction see Ref. 26.}
\REF\collins{J.C.  Collins, {\it Renormalization} (Cambridge U. Press,
Cambridge
             1984).}
\REF\itzykson{C. Itzykson and J.-B. Zuber, {\it Quantum Field Theory}
              (McGraw-Hill, New York, 1980).}
\REF\perispeccei{R.D. Peccei and S. Peris, \prd44 (1991) 809.}
\REF\chiral{See, for example,  Refs. \georgi, \pich, 42, 49, 64, 70, 36, 37.}
\REF\wess{ J. Wess and B. Zumino, \plb37 (1971) 95.
           E. Witten, \npb223 (1983) 422.}
\REF\farhidhoker{E. D'Hoker and E. Farhi, \npb248 (1984) 59,
                 \ibid 77. See also T. Sterling and M. Veltman
                 \npb189 (1981) 557.}
\REF\yao{H. Steeger \etal, \prl\ {\bf59} (1987) 385.
         G.-L. Lin \etal, \prd44 (1991) 2139; University of Michigan
         report UM-TH-93-05 (unpublished).}
\REF\colemanN{S. Coleman, lectures presented at the
              {\it 1979 Int. School of Subnuclear Physics, Pointlike
              Structures Inside and Outside Hadrons,} Erice, Italy,
              Jul 31-Aug 10, 1979.}
\REF\gasser{J. Gasser \etal, \npb307 (1988) 779.}
\REF\appelquist{T. Appelquist, \prd22 (1980) 200.}
\REF\bernard{C. Bernard, \prd23 (1981) 425.}
\REF\burges{C.J.C.  Burges and H.J. Schnitzer, \npb228 (1983) 464.}
\REF\keung{C.N. Leung \etal, \zphys31 (1986) 433.}
\REF\bw{W. B\"uchmuller and D. Wyler, \npb268
        (1986) 621; see also W. B\"uchmuller \etal, \plb197
         (1987) 379.}
\REF\peskin{M. Peskin and T. Takeuchi, \prl65 (1990) 964.
            G. Altarelli and R. Barbieri, \plb253 (1991) 161.
            B. Lynn \etal, in {\it Physics at LEP}, CERN
            Yellow report 86-02.}
\REF\coleman{S. Coleman \etal, \pr177 (1969) 2239.
             C.G. Callan \etal, \pr177 (1969) 2247.}
\REF\sikivie{P. Sikivie \etal, \npb182 (1981) 529.}
\REF\peccei{R.D. Peccei and X. Zhang, \npb337 (1990) 269.}
\REF\appelquistwu{T. Appelquist and G.-H. Wu, \prd48 (1993) 3235.}
\REF\dougangolden{M.J. Dugan and M. Golden, report
                  BUHEP-93-14,HUTP-93/A016, hep-ph/9306265 (unpublished).}
\REF\chivukula{S. Chivukula \etal, \prd47 (1993) 2930.}
\REF\appelquistterning{T. Appelquist and J. Terning, \prd47 (1993) 3075.}
\REF\georgibook{H. Georgi, {\it Weak Interactions and Modern
                Particle Theory} (Benjamin/Cummings, Menlo Park, CA, 1984).}
\REF\georgimanohar{H. Georgi and A. Manohar, \npb234 (1984) 189.}
\REF\gounarisi{G. Gounaris \etal, report PM 93/26 (unpublished).}
\REF\eichten{E. Eichten \etal, \prl50, 811 (1983).}
\REF\ruckl{R. R\"uckl, \plb129 (1983) 363; \npb234 (1984) 91.
           See also R.J. Cashmore \etal, {\it Phys. Rep.} 122 (1985) 275.}
\REF\ehlq{ E. Eichten \etal, {\it Rev. Mod. Phys.} { \bf56} (1984) 579.}
\REF\djouadi{A. Djouadi \etal, in {\it Physics and Experiments
             with Linear  Colliders}, R. Orava \etal, editors,
             Saariselk\"a, Finalnd, 9-14 Sept. 1991 (World Scientific,
             Singapore, 1992).}
\REF\doncheski{K.A. Doncheski,\zphys52 (1991) 527.}
\REF\barbiellini{G. Barbiellini \etal, in {\it Physics at LEP}, edited
                 by J. Ellis and R. Peccei, CERN Yellow report 86-02.}
\REF\einhornandwudka{M.B. Einhorn and J. Wudka, \prd39 (1989) 2758, \ibid
                     47 (1993) 5029.}
\REF\wudka{J. Wudka, {\it J. Phys.} {\bf A25} (1992) 2945.}
\REF\arztii{C.Arzt \etal, in preparation.}
\REF\veltmaniruv{M. Veltman, {\it Acta Phys. Polon.}{\bf B12} (1981) 437.}
\REF\rajaraman{R. Rajaraman, {\it Solitons and Instantons}
               (North Hollan Pubishing Co., Amsterdam, Holland, 1982).}
\REF\arztiii{C. Arzt, Univ. of Michigan report  UM-TH-92-28 (unpublished)
             (Bulletin Board: hep-ph@xxx.lanl.gov - 9304230).}
\REF\golden{M.Golden and L. Randall, \npb361 (1991) 3.}
\REF\escribano{R. Escribano and E. Mass\'o, \plb301 (1993) 419.}
\REF\ahn{C. Ahn \etal, \npb309 (1988) 221.}
\REF\chanowitz{M. Chanowitz \etal, \npb153 (1979) 402.}
\REF\changeogol{M. Chanowitz \etal, \prd36 (1987) 1490.}
\REF\kennedy{W.E. Caswell and A.D. Kennedy \prd25 (1981) 392.}
\REF\longhitano{A. Longhitano, \npb188 (1981) 118.}
\REF\arzti{C.Arzt \etal, U.C. Riverside report UCRHEP-T98 and \prd\ (to
appear).}
\REF\roy{D. Choudhury \etal,  Tata Institute report TIFR-TH/93-08
         (unpublished).}
\REF\marciano{T. Kinoshita and W. Marciano in {\it Quantum Electrodynamics},
              T. Kinoshita Ed. (World Scientific Singapore, 1990).}
\REF\gunionetal{J.F. Gunion \etal, {\it The Higgs Hunter's Guide},
                (Addison-Wesley, Redwood City, CA, 1990).}
\REF\perez{M.A. Perez \etal, in preparation.}
\REF\nielsen{D. Forster and H.B. Nielsen, \plb92 (1980) 128.}
\REF\anom{J. Minn \etal, \prd35 (1987) 1872.}
\REF\partdatbook{K. Hikasa \etal (Review of Particle
                 Properties), \prd45 (1992) 1.}
\REF\ags{V.W. Hughes, AIP conference proceedings no. 187, 326 (1989.)
         M. May, AIP conference proceedings no. 176, 1168 (1988).}
\REF\cleo{E. Thorndike, CLEO collaboration, talk given at the
          {\it 1993 Meeting of the American Physical Society},
          Washington, D.C., April, 1993.}
\REF\hemackellar{X.-G. He and B. McKellar Univ. of Melbourne report
                 UM-P-93-52 (unpublished) (Bulletin Board:
                 hep-ph@xxx.lanl.gov - 9309228).}
\REF\peterson{K.A. Peterson, \plb282 (1992) 207.}
\REF\dawson{S. Dawson and G. Valencia Fermilab report FRMILAB-PUB-93/218-T
             and \prd\ (to appear).}
\REF\burgesschnitzer{C.J.C.  Burges and H.J. Schnitzer, \plb134 (1984) 329.}
\REF\wudkahera{J. Wudka, Univ. of California Riverside report T113 and
               \prd (to appear).}
\REF\helbig{T. Helbig and H. Spiesberger, \npb373 (1992) 73.}
\REF\kim{C.S. Kim \etal, YUMS 93-15, SNUTP 93-44.}
\REF\hagiwaraii{K. Hagiwara \etal, \prd48 (1993) 2182.}
\REF\kneur{J.-L. Kneur \etal, \plb262 (1991), 93.}
\REF\greenstein{B Grinstein and M. Wise, \plb265 (1991) 326.}
\REF\holdomi{B. Holdom, \plb259 (1991) 329.}
\REF\ellis{J. Ellis \etal \plb292 427 1992.}
\REF\wudkamerida{J. Wudka, lectures presented at the {\it IV Workshop
                 on Particles and Fields,} M\'erida,
                 Yucat\'an M\'exico, Oct 25-29, (1993).}
\REF\hernandez{P. Hern\'andez and F.J. Vegas,  \plb307 (1993) 116.}
\REF\wainer{N. Wainer, talk presented at
             the {\it 1993 Aspen Winter Conference on Elementary Particle
             Physics}, Aspen Center for Physics, January 10--16, 1993.}
\REF\hagiwarazep{U. Baur and E.L. Berger, \prd41 (1990)
                 1476. K. Hagiwara \etal, \ibid\ pp 2113.}
\REF\bauri{U. Baur \etal, in
           proceedings of the {\it  Workshop on Physics at Current
           Accelerators and the Supercollider}, Aragonne, IL, 2-5 Jun 1993.}
\REF\ellison{A.L. Spadafora: the D\O\ Collaboration, to appear in
             {\it $ 9^{ th } $ Topical Workshop on $ \bar p p $ Collider
             Physics}, Tsukuba, Japan, Oct. 18-22, 1993.}
\REF\bilenky{M. Bileny \etal, Univ. of Bielefeld report BI-TP-92/44
             (unpublished).}
\REF\holdomii{B. Holdom, \plb258 (1991) 156.}
\REF\boudjema{F. Boudjema, in {\it 2nd International Workshop on Physics
              and Experiments with Linear $ e^+ e^-$ Colliders},
              Waikoloa, HI, 26-30 Apr. 1993.
              (Bulletin Board: hep-ph@xxx.lanl.gov - 9308343).}%
\REF\falk{A.F. Falk \etal, \npb365 (1991) 523.}
\REF\diakonos{F.K. Diakonos, \etal, CERN report CERN-TH.6753/92
              (unpublished).}
\REF\laser{F.R. Arutyunian and V.A. Tumanian, {\it Phys. Lett.}4 \
           (1963) 176 ;
           R.H. Milburn, \prl10 (1963) 75.
           I.F. Ginzburg \etal, \npb228 (1983) 285.
           I.F. Ginzburg \etal, {\it Nucl. Instrum. \& Methods}205
           (1983) 47;\ibid 219 (1984) 5. {\it Nucl. Instrum. \& Methods}
           A294 (1990) 72.}
\REF\likhoded{A.A. Likhoded \etal, report IC/93/288 (unpublished)
              (Bulletin Board: hep-ph@xxx.lanl.gov - 9309322).}
\REF\barklowi{T. Barklow, in {\it Physics and Experiments with Linear
              Colliders,} R. Orava \etal, editors,
              Saariselk\"a, Finalnd, 9-14 Sept. 1991 (World Scientific,
              Singapore, 1992).}
\REF\yehudai{E. Yehuday,\prd41 (1990) 33; \ibid\ {\bf D}44 (1991).}%
\REF\grossekneterschildknecht{C. Grosse-Knetter and D. Schildknecht,
                              \plb302 (1993) 309.}
\REF\barklow{T. Barklow, talk presented at
             the {\it 1993 Aspen Winter Conference on Elementary Particle
             Physics}, Aspen Center for Physics, January 10--16, 1993.}
\REF\gonzalez{O.J.P. Eboli \etal, Univ. of Wisonsin report MAD-PH-764
              (unpublished). (Bulletin Board: hep-ph@xxx.lanl.gov
              - 9306306).}
\REF\pereztoscano{M.A. Perez and J. Toscano, \plb289 (1992) 381.}
\REF\konig{H. K\"oing, \prd45 (1992) 1575.}
\REF\martypion{M.B. Einhorn, Univ. of Michigan report UM-TH-93-18
               (unpublished) (Bulletin Board: hep-ph@xxx.lanl.gov -
               9308254).}
\REF\fujikawa{K. Fujikawa, \prl {\bf42} (1979) 1195;
              \prd21 (1980) 2848 (1980).}



\chapter{Introduction.}

When constructing models aimed at describing physics behind
a certain set of phenomena the resulting formalism
is understood to have, except in the most ambitious cases
(such as superstrings) a limited range of validity.
For example, hydrodynamics
is very successful in describing liquid motion at scales much
larger than the typical atomic size. Below such scale the model
describing the behaviour of this system changes dramatically as the
dynamical variables are different.
 Of course, barring technical difficulties, the
hydrodynamic description of a fluid can be obtained from the
microscopic one by defining the appropriate macroscopic observables
and deducing the quantum-average equations of motion. Going a step
further one can imagine deducing from the quantum theory the
$ O ( \hbar ) $ corrections to the Navier-Stokes equation. Then
we would find that, as we consider smaller and smaller distances,
these corrections will become increasingly important, to the point
that when we reach distances comparable to the atomic scale the
whole series in $ \hbar $ must be included in order to describe the
system.

In the realm of high energy physics there is a direct parallel with the
above observations. When the description of a certain set of
phenomena in terms of a model
is desired, the first ingredient in the construction of the
model is the determination of the relevant particles, corresponding to
the dynamical variables of the theory. Then, often
based on experimental evidence, the symmetries obeyed by the model
are specified. Finally the most general (local) lagrangian
is constructed containing the
fields corresponding to the above particles obeying the said
symmetries is constructed.

To determine the physical content of an effective theory it is
necessary to understand how it is generated
(for a review see Ref.~\polchinsky). Suppose we have
a model from which we wish to derive a description of phenomena
at energies below a certain scale $ \Lambda $. This description is
formally obtained from an effective action $ S^\Lambda \lowti{eff} $
whose 1PI functions
are generated by loops containing only excitations of energy
$ \ge \Lambda $; equivalently $ S^\Lambda \lowti{eff} $ is derived
from a functional integral  by integrating over all Fourier
components of energy $ \ge \Lambda $. The resulting expression
will be a non-local function of the low energy fields
(\ie, those whose Fourier components correspond to energies $ < \Lambda $).
Note that a simple description of this effective action might require
the low energy fields to be non-linear functions of the original
fields. This is the case, for example, in QCD where the low energy
physics is best described in terms of composite quark operators
corresponding to the various mesons and baryons.

The effective action
can be expanded in an infinite series of local
(effective) operators $ \ocal_i $ containing only low energy fields,
with  $ \Lambda $-dependent coefficients\refmark{\polchinsky,\wilson}
$$ S^\Lambda \lowti{eff} =
\int d^4x \; \leff = \int d^4x \; \sum_i \alpha_i ( \Lambda ) \ocal_i .
\eqn\defleff $$
which defines the effective lagrangian $\leff$.
The effective lagrangian comprises all
virtual heavy physics effects; as long as we remain within
its realm of applicability (at scales below $ \Lambda $), it provides
a unitary, consistent and complete description of all low energy
phenomena;~\refmark{\weinberg} in particular all the Green functions
of the heavy theory at low energies can be derived from \defleff.

The {\it form}
of the effective lagrangian in \defleff\ is independent of the
model from which it is derived. It then follows that one can
parametrize all possible heavy physics effects by an
expression of this type, where, as
mentioned above, the $ \ocal_i $ are only constrained
by the symmetries of the low energy physics.
This expression for $ \leff $ will provide
a model-independent, consistent, complete and unitary description of
heavy physics effects which respects all low energy constraints
required by the corresponding symmetries; for
example, Green functions including heavy physics
effects are parametrized in this
manner independently of the details of the heavy physics.
It is because of
its generality that the effective lagrangian approach is ideally
suited for studying possible effects of physics beyond the \sm.

The effective lagrangian contains
several length scales; some of these are identified with the
properties of the particles under consideration, such as the
corresponding Compton wavelengths. Other scales are not so
associated, and are related to the physics underlying this theory.
The model also possesses a number of  coupling constants
that are determined using experimental data. It is expected (on
\ae sthetic grounds) that a fundamental theory will have a very
small number of parameters; but a more modest model, aiming at
understanding a limited set of processes, can have a large number of
such couplings, in some cases an infinite number. These
models have less predictability than an ideal fundamental theory, but,
even in the case where there are an infinite number of couplings, the
models do have predictive power, as we will see below.

I have not mentioned renormalizability  in connection with
the above models since they are not
supposed to be a good description of nature above a certain energy
scale, and so  ultra-violet divergences are absent. Since
renormalizability is not imposed, the lagrangians will
in general contain an
infinite set of local operators constructed out of the dynamical
variables of the theory and satisfying the symmetries of the model.
Each such operator will be multiplied by
an undetermined coupling constant (dimensionless or
dimensionful) whose precise value is determined by
the unknown heavy physics. These constants parametrize
all heavy physics effects at low energies.

The presence of an infinite set
of couplings leads, via ``standard arguments'', to a complete
lack of predictability, which apparently implies that this type of model
has little practical interest. This is in fact not so:
we will see that in these models one can define a hierarchy in the
operators present in the lagrangian; to any order in this hierarchy
there is a finite set of couplings to consider, and so the model can
produce non-trivial predictions. Within the range of applicability
of the model the contributions to any process is dominated
by the lowest-order operators. One can also estimate the
corrections to these predictions produced by the higher order
operators; these models have predictive power and also
provide information about their range of applicability.

In the following I will use the label ``effective theories'' to denote
the set of models
described above. I will show that,
despite their not being ``fundamental'', they are extremely useful
in understanding physics in a limited range of
scales. It is the
purpose of this paper to present a review of this type of
models, to describe their properties and advertise their usefulness.
Many of the ideas presented in this review have appeared
elsewhere.~\refmark{\weinberg,
\polchinsky,\georgi,\martyjose,\derujula,\londonburgessi} One
purpose of this review is to present these ideas and summarize
the conclusions that have been drawn from them.

The contents of this paper are the following. In the next section
the symmetries local and global of effective models  are discussed, with
special attention paid to the issue of gauge invariance.
Section 3 presents the construction of the effective
lagrangian for electroweak interactions, while section 4
describes the determination of the order of magnitude of the
(unknown) coefficients which appear in it.
Section 5 discusses the use of the equations of motion to reduce
the number of terms in the effective lagrangian; section 6 presents
a brief description of the problems associated with unitarity within
the context of effective theories. Section 7 illustrates the use
of effective lagrangians in calculating radiative corrections.
Applications of the effective lagrangian parametrization are presented
in section 8 where the limits on various coefficients derived
from known data are summarized and, also, the expected sensitivity
of future colliders to the new interactions is presented.
Conclusions are given in section 9 and several useful expressions
and a simple calculation are presented in the appendices.

The notation used in this paper when referring to the \sm\
fields follows that of [\bw]; it is summarized here for
convenience. The fields are
$$ \eqalign{
W_\mu^I &: \su2_L \ \hbox{{gauge field}} ; \cr
B_\mu   &: \ui_Y \ \hbox{{gauge field}} ; \cr
G_\mu^A &: \su3_c \ \hbox{{ gauge field}} ; \cr
\phi &: \hbox{scalar doublet} ; \cr
\ell_l &: \hbox{left-handed lepton doublet with charged lepton}
l, \ l = e, \mu ,\tau ; \cr
l_R &: \hbox{right-handed charged lepton,} \ l= e , \mu , \tau ; \cr
q &: \hbox{left-handed quark doublet} ; \cr
u_R , \ d_R &: \hbox{right-handed quark fields, up and down type
respectively } ; \cr } \eqn\convi $$
The gauge field curvatures are
$$ \eqalign{ W_{ \mu \nu }^I &= \partial_\mu W_\nu^I - \partial_\nu W_\mu^I
+ g \epsilon_{ I J K } W_\mu^J W_\nu^K ; \cr
B_{ \mu \nu } &= \partial_\mu B_\nu - \partial_\nu B_\mu ; \cr
G_{\mu \nu}^A &= \partial_\mu G_\nu^A - \partial_\nu G_\mu^A + g_s
f_{ A B C } G_\mu^B G_\nu^C . \cr } \eqn\convii
$$ The corresponding gauge couplings are $g$ for $ \su2 $, $ g' $ for
$ \ui $ and $ g_s $ for $ \su3 $. The structure constants for $ \su3 $
are denoted by $ f_{A B C } $.

The covariant derivative for a colorless $ \su2$ doublet of hypercharge
$Y$ is $$ D_\mu = \partial_\mu - i g \half \sigma_I W_\mu^I
- i g' Y B_\mu . \eqn\conviii $$ The hypercharge assignments are
$ Y ( \phi ) = 1/2 $,
$ Y ( \ell_l ) = -1/2 $,
$ Y ( l_R ) = -1 $,
$ Y ( q ) = 1/6 $,
$ Y ( u_R ) = 2/3 $,
$ Y ( d_R ) = -1/3 $.
I will also need the matrix $ \epsilon = i \sigma_2 $ and the field
$ \tilde \phi = \epsilon \phi^* $.

\def\con{(the notation is given in Eqs.~\convi-\conviii\ and the comments
following them.)}

There are several topics which will not be discussed in this review.
I will not cover the BESS model,~\refmark{\casalbuoni} for which I refer
the reader to the literature.
No mention will be made of the application of effective lagrangians
to low energy strong interactions. This has been covered in many
excellent publications (see Ref.~\gl; for a recent review see Ref.~\pich).
On the weak interactions area there are many publications dealing
with specific models and their low energy phenomenology (see
for example the extensive review in Ref.~\langacker\ and references therein),
this approach will not be followed since, by definition, is model
specific.

Several authors~\refmark{\herrero,\bagger,\baggerdawson} have also considered
a mixed approach where the effective models are studied beyond their
UV cutoff; this generates, among other problems, violations of
tree level unitarity.~\footnote{a}{See section 6 for a discussion
on this point} To fix this problem several methods of ``unitarizing''
the models are used (see the above references as
well as Ref.~\chanowitzandgaillard; for a recent review see
Ref.~\bargerandphillips). This approach involves specific
assumptions regarding the new physics and to this extent runs
counter to the philosophy of using an effective lagrangian parametrization.
This approach will not be presented here. I refer the reader to the
above references for a thorough discussion.

There are also several publications where the coefficients of the
effective lagrangian terms are taken to be form
factors.~\refmark{\bauzeppenfeld} This will also not be covered in this review
since
it is hard to translate the results obtained using the
form-factor approach to the language used in this paper. I refer
the interested reader to the literature.

\chapter{Symmetries.}

One of the most important ingredients in the construction of effective
lagrangians, denoted by $ \leff $,
is the determination of the symmetries respected by the models.

\section{{\bfit{Global symmetries.}}}

Global symmetries such as those associated with lepton or baryon number
conservation, are imposed based on  experimental evidence.
The requirement that $ \leff $ satisfies them just imposes an
added restriction which reduces the number of allowed operators.
Should  experiments demonstrate a violation of these symmetries,
one can always introduce effective interactions which
violate them. These interactions will then be associated with the scale
at which violations of the global
symmetry are generated. For very small deviations (as, for example, for
baryon number violation) such a scale will be very large compared to
the typical scale of the low energy physics. Thus, for example,
there is no fundamental theoretical principle which
forbids the introduction, in an electroweak effective lagrangian,
of the term $$ \inv{ \Lambda }
\bar e_R \; \sigma_{\mu \nu } \; \mu_R B^{ \mu \nu } \eqn\eq $$
\con. This interaction generates
the decay $ \mu \rightarrow e \gamma $ with a width
$ \propto 1/ \Lambda^2 $. The fact that this process has not been seen
merely implies that $ \Lambda $ is very large.

\section{{\bfit{Local symmetries.}}}

Many phenomenological models considered in the literature
which describe physics beyond the \sm\ effects are not
gauge invariant \refmark{\hagiwara}, and because of this have been
strongly criticized.~\refmark{\derujula}
It was then pointed out~\refmark{\goldenpriv, \burgessandlondon}
that {\it any} effective lagrangian can be understood as the unitary
gauge limit of a gauge invariant $ \leff $. In this section I will
discuss these issues. Other related considerations
concerning gauge invariance and radiative corrections will be presented
in section  7.3.

The process by
which a lagrangian is
made gauge invariant is based on
a trick originally invented by St\"uckelberg (see Refs.~\gates\ and
\delbourgo).
The idea is the following: suppose we have a lagrangian $ \lcal $ depending
on some (real)
vector fields $ W_\mu^n , \  n = 1, 2, \cdots , N  $, together
with some other fields, which I'll denote
by $ \chi $. Choose then a (Lie) group  $ \gcal $
of dimension $ D \ge N $, and add $ D-N $ auxiliary (real) vector
fields $ \bar W^n_\mu , \ n = N+1 , \cdots , D $ assumed to be
non-interacting (if desired, the masses for these fields can be taken
 $ \ge \Lambda $); this modification of the effective lagrangian does
not affect any low energy observable.
Henceforth I will drop the over-bar in these extra fields;
the lagrangian (with  the new fields) will be denoted by $ \lcal
( W ; \chi ) $.

Using the above vector fields, and the (antihermitian)
generators of $ \gcal $, denoted by
$ \{ T^n \} $ (in any \irrep), normalized such that
$ \tr \{ T^n T^m \} = - \delta_{ n m } $,
I can construct the derivative operator
$$ D_\mu = \partial_\mu + \sum_{ n = 1 } ^D T^n W^n_\mu . \eqn\eq $$

Next I introduce a unitary auxiliary field $U$ which, under an
infinitesimal $ \gcal $ transformation, behaves as $ \delta U = \sum_n
\epsilon_n T^n \cdot U $.
Finally I define the objects
$$ \wcal_\mu^n = - \tr \left( T^n \; U^\dagger D_\mu U \right) . \eqn\eq $$

With these definitions I now assume that the $ W^n $ are in fact
gauge fields with all the corresponding properties.
Then the $ \wcal^n $ are gauge invariant objects which satisfy $$ \left.
\wcal^n_\mu \right|_{ U = 1 } = W_\mu^n . \eqn\eq $$ It follows that
$$ \lcal ( W ; \chi ) = \left.
\lcal ( \wcal ; \chi ) \right|_{ U = 1 } , \eqn\eq $$ which means that the
original lagrangian is the unitary gauge limit of a gauge invariant theory.
The fields $ \chi $ are assumed to be gauge invariant.

The price paid in rendering a model
gauge invariant is, once $ \gcal $ is chosen,
the (possible) introduction of the extra vector fields $ \bar W $ and the
auxiliary field $U$.
Note that even if no extra vector fields are introduced, there
is always a group of dimension $N$: $ \ui^N $. Therefore the above
procedure can also be carried out with the original vector fields
as gauge fields, though this restricts the allowed choices for $ \gcal $.

On the basis of the above construction it has been
argued~\refmark{\burgessandlondon}
that gauge invariance has no content. After all, any theory
can be considered as a gauge invariant theory, albeit in the
unitary gauge. I disagree with this conclusion. Obvious drawbacks
of the procedure on which this statement is based are
the arbitrariness in the choice of $ \gcal $, and the fact that
the construction in general disallows a
linear realization of the gauge symmetry~\footnote{b}{ The exception
corresponds to the case where $ \gcal $ is a product of $ \ui $
factors.}(see section 3.2 for further discussion).

Consider for example the \sm\ in the unitary
gauge. Taking the corresponding lagrangian
as the terms of dimension $ \le4 $ of an effective
theory, the above trick
can be used to render it an $ \su2 \times \ui $ {\it or} a $ \ui^4 $
gauge theory.
In the latter case experimental results would suggest
a number of relations
among the coefficients of the operators allowed by the symmetry,
such relations would have
no fundamental justification. If on the other hand we assume that the
above group is $ \su2 \times \ui $ these relations between the coefficients
would be {\it predicted} by this model, considerably adding to its
credibility.

More important is the fact that the non-gauge fields $ \chi $ are assumed
to be gauge singlets. Thus, for example, the relationships between the
$Z$, $W$ and photon couplings to the fermions would, in this approach,
be a mere accident (it is certainly
possible to impose the \sm\ relations among
these couplings, but there would be no justification for this choice).
If, on the other hand, the \sm\ local symmetries and
the representations carried by the fields are chosen, then these
relations are again a success of the model.

These arguments support the claim that gauge invariance
does have non-trivial content, not in the abstract sense (since indeed
any model can be made gauge invariant
under, in general, many gauge groups),
but in the practical sense: when it is stated that a given theory is
gauge invariant with group $ \gcal $
and the corresponding transformation properties of the fields
are specified, the structure of the model, together
with the number of unknown parameters, is largely fixed. This {\it can} be
tested by experiment. If gauge invariance had no content these conditions
would be naturally predicted irrespective of $ \gcal $.

These considerations are of importance when doing loop calculations:
the sensible approach is then to choose
a renormalizable gauge and this requires the addition of
a gauge fixing lagrangian and the corresponding Fadeev-Popov
ghost terms. Thus the lagrangian used in the actual calculations
will be quite different from the original one; the results might
also vary due to the singular nature of the unitary
gauge.~\refmark{\grossekneterandkogerler}

The effects of gauge invariance at tree
level calculations are also important. The fact that
the effective operators are gauge invariant implies that certain
vertices with a different number of legs are closely related. Consider
for example, for the \sm\ gauge symmetry,
the operator $ \phi^\dagger \sigma_I \phi B^{ \mu \nu }
W^I_{ \mu \nu } $ \con.
This operator affects the oblique $S$ parameter (section 8), but it
also modifies the anomalous magnetic moment of the fermions, $W$
pair production, and Higgs production at colliders; and all these
contributions appear with a single coupling. There are of course
many more operators contributing to these processes, but the
requirement of gauge invariance does significantly restrict their
number.

\section{{\bfit{Comments.}}}

Throughout this paper I will assume that the local gauge symmetry is
 $ \su2_L \times \ui_Y $ as in the \sm. The global symmetries will
be taken to be the ones respected by the \sm.

These choices are of course not mandatory. As mentioned above,
effective operators that violate the global
symmetries are easily constructed (though
experimental constraint require the associated scale(s) to be very large).
It is also possible to choose a different
gauge group. The reasonable approach is then to select a group $ \gcal $
which contains the \sm\ gauge group. This is
investigated in Ref.~\frere\ for the case
$ \gcal = \su2_L \times \ui \times \ui' $. The modification of the gauge
group implies in general that the low energy sector is richer than in
the \sm; for the example considered there is an
additional scalar singlet (under $ \su2_L $) and a neutral gauge
boson $Z'$, aside from the right handed neutrino fields  required
to cancel anomalies. The presence of an enlarged symmetry
imposes further restrictions on the allowed operators; on the other
hand the increased particle content allows for more operators to
be constructed. For the model studied in Ref.~\frere, the second of these
opposing tendencies dominates:
the enlarged low energy spectrum generates a significantly larger
number of operators. The phenomenological consequences of this
are presented in section 8.7.

\chapter{Effective Lagrangians.}

Following the procedure outlined in the introduction one can
generate an effective lagrangian by constructing, using the
relevant fields, all local operators satisfying the required
symmetries. The effective lagrangian $ \leff $ is then equal
to the infinite sum of these operators multiplied by unknown
(dimensionful or dimensionless) coefficients. Such an object
is understood to be the low energy limit of a theory whose
presence will become apparent at energies of the order of a scale
$ \Lambda $. All dimensionful parameters will be proportional to $
\Lambda $ to the appropriate power unless they are fine tuned or are
protected by a symmetry; an example of the second possibility is the
use of chiral symmetry to insure naturally light fermion
masses.~\refmark{\thooft} For an excellent discussion of these issues see
Ref.~\polchinsky.

It is of course
possible for different kinds of new physics to be generated at
different scales. For example, the CP violating operators could
be generated by physics whose structure becomes apparent at
a scale $ \Lambda' $. If this is the case then $ \Lambda $ will
denote the smallest of such scales. It must be kept in mind,
however, that some operators can have suppression factors
of the type $ ( \Lambda / \Lambda' )^n $ for some integer $n$.

{}From $ \leff $ we can extract those terms containing operators
of (canonical) dimension $ \le 4 $. The corresponding object
will be denoted $ \liv $;  $ \leff - \liv  $ is an infinite series
of local operators each suppressed by the appropriate power of
$ \Lambda $. Some important characteristics of $ \leff $
will depend on whether $ \liv $ is renormalizable or not.

If $ \liv $ is renormalizable then the conditions for the decoupling
theorem~\refmark{\decoupling, \collins} are satisfied. This
insures that the contributions of the
terms in $ \leff - \liv $ to any observable appear as a power
series in $ 1/ \Lambda $. In particular all effects from the dimension $
> 4 $ operators vanish as $ \Lambda \rightarrow
\infty $.~\footnote{c}{There are
some contributions to Green's functions that grow with $ \Lambda $, but
they can all be absorbed in a renormalization of the
parameters in $ \liv $;
these contributions are important when considering the naturality of
the model.} The situation where $ \liv $ is renormalizable
will be labelled the {\it \dec}.

The \dec\ is typically realized
when $ \leff $ is obtained from the heavy physics lagrangian
by letting a dimensionful parameter become very large. For example,
in the limit in which the Higgs \vev\ $v$ in the \sm\ becomes very large
(or equivalently when we are concerned with processes at energies
significantly
below the $W$ and $Z$ vector boson masses) we obtain QED together
with an infinite tower of operators suppressed by inverse powers of $v$,
as is the case for the four-fermi interactions.~\refmark{\itzykson}

If $ \liv $ is not renormalizable the decoupling theorem
does not hold. The divergences generated by
$ \liv $ cannot be absorbed in its own coefficients,
and require terms
from $ \leff - \liv $ in order to  renormalize the
theory.
In many phenomenologically interesting cases this situation is associated
with a derivative expansion: terms in $\leff$ are collected
according to the number of derivatives they carry.~\footnote{d}{When
fermions are included a slight modification is required,
see below.} In this case we will see that there are effects
which do not vanish
as $ \Lambda \rightarrow \infty $ despite being associated with the
heavy physics (a simple example is the contribution to the oblique
parameter $S$ form  a heavy generation~\refmark\perispeccei).
This situation will be denoted the {\it \ndec}.

In contrast to the \dec, the \ndec\ is typically
realized when $ \leff$ is obtained by letting a dimensionless parameter
become large. The reason for this difference is that the constraint
imposed by letting a dimensionful  parameter become infinite consists in
setting
the corresponding field to zero. In contrast, taking a dimensionless
parameter to  infinity, will generate non-linear
interactions among the fields.

A familiar example of the \ndec\ is obtained when the scalar self-coupling
constant in the \sm\ is assumed to be large. In this case
we are led to a scalar sector without a Higgs particle: a non-linear
(gauged) sigma model~\refmark{\chiral} (see section 3.2 below).

Another example for the \ndec\
is generated when one or more Yukawa couplings in the \sm\
are taken to be very large. When both the
Yukawa couplings of a quark doublet are taken to infinity,
the low energy effective lagrangian
will include a tower of non-linear scalar interactions. Among
these the appropriate Wess-Zumino lagrangian~\refmark{\wess} is
generated which insures
that the resulting (effective) lagrangian is anomaly
free.~\refmark{\farhidhoker}
A more phenomenologically relevant investigation corresponds to the
case where only one member of a quark doublet, the top,
is assumed to have a large
Yukawa coupling.~\refmark{\yao} In this case the constraint on the
top--bottom quark doublet is, at tree level,
$$ t_L = - \left( { \phi_+ \over \phi_0 } \right) b_L ,
\qquad t_R = 0 \eqn\eq $$
where $ \phi_{ + , 0 } $ are the components of the scalar doublet:
$ \phi = { \phi_+ \choose \phi_0 } $.
The full
set of operators present in $ \leff $ including tree and one loop
top effects can be found in Ref.~\yao. In this case, since the underlying
theory is described by the \sm\ lagrangian, the coefficients of all
the effective operators can be determined explicitly.

Further insight into these two cases
can be gleaned by using a simple power counting argument.
Consider
a theory whose vertices are labelled by an index $n$; the vertex of
type $n$ will have $ b_n $ bosonic lines, $ f_n $ fermionic lines
and $ d_n $ derivatives. The dimension of an $L$ loop
integral (excluding dimensionful couplings) with
$ I_B $ internal boson lines, $ I_F $ internal fermionic lines
and $ V_n $ vertices of type $n$, is
$ D = 4 L - 2 I_B - I_F  + \sum V_n d_n $. In order to renormalize
the divergences appearing in this graph
we need local counterterms of dimension $ \le D $ (whenever $ D \ge  0 $).
Using the relations
$$ \sum b_n V_n = 2 I_B + E_B , \quad
\sum f_n V_n = 2 I_F + E_F , \quad
L = 1 + I_B + I_F - \sum V_n , \eqn\eq $$
($ E_B $ and $E_F$ denote the number of external bosonic
or fermionic lines respectively), the expression for $D$ implies
$$ S ( u ) = ( 4 - u ) L + \sum_n V_n \; s_n ( u ) , \eqn\indeq $$
where $$ \eqalign{ s_n ( u ) &= d_n + \left( \half u - 1 \right) b_n
+ \half ( u - 1 ) f_n - u ; \cr
S ( u ) &= D + \left( \half u - 1 \right) E_B
+ \half ( u - 1 ) E_F - u ; \cr} \eqn\indeq $$
and  $u$ is an arbitrary (real) parameter.
I will call $ s_n ( u ) $ the {\it``index''} of the vertex of type
$n$.

This treatment of power counting, though somewhat unconventional
due to the presence of $u$, has the advantage of unifying
several interesting situations to be discussed next. For example
recall that power counting in the chiral approach
to the strong interactions\refmark{\pich}\ differs from the
power counting used in evaluating ultra-violet divergences.
Both these cases can be dealt with by appropriate choices of $u$.

Equation~\indeq\ implies that the $L$ loop graphs under consideration
generate vertices of index $ S ( u ) $. This, in its turn, implies
that the natural size of the
coefficients of the vertices with index $ S ( u ) $ in $ \leff $
are of the same order as those obtained from these
graphs.

If the parameter $u$ is chosen so that $ S ( u ) \ge s_n ( u ) $
in \indeq\ (for any graph and for any $L$), and the terms in $\leff $
are ordered according to their index, that is $$
\leff = \sum_{ i = 0 }^\infty \lcal_i \qquad
\hbox{index} \left( \lcal_{ i + 1 } \right) >
\hbox{index} \left( \lcal_i \right) ; \eqn\leffordering $$
then \indeq\ also insures that this hierarchy is consistent
with the loop expansion: the index of a graph is never smaller than
the indices of its vertices.

For example, in the \ndec\ below, we will find it
useful to choose $ u = 2 $. Then the index is independent of
the number of bosonic legs: $ s_n ( 2 ) = d_n + f_n/2 - 2 $
and $ S ( 2 ) = 2 L + \sum V_n s_n ( 2 ) $, which will
be used below. This relation implies
the following. Suppose  $ \lcal_0 $ in \leffordering\
contains terms with
$ s_n ( 2 ) =0 $,~\footnote{e}{ The terms with $ d_n = 0 $ and $ f_n = 2 $,
corresponding to the fermion mass terms in $ \leff $, require special
considerations, see below.} then the one
loop graphs generate terms with index $ s_n = 2 $. The two loop graphs with
vertices in $ \lcal_0 $ together with the one loop graphs containing one
index $ s_n = 2 $ vertex (generated by $ \lcal_1$)
generate vertices with index $ s_n = 4 $, etc.
For the purely bosonic terms in $ \leff $
the ordering \leffordering\
is equivalent to the derivative expansion familiar
from the chiral lagrangian approach to low energy
strong interactions.~\refmark{\gl, \gasser, \pich, \appelquist, \bernard}

Choosing now $ u = 4 $
gives $ S ( 4 ) = \sum V_n s_n ( 4 ) $ where $ s_n ( 4 ) + 4 $ is just
the canonical dimension of the vertex of type $n$; this choice will be
relevant for the \dec. In this case the terms in $ \leff $ are
ordered according to their dimension or, for the terms in $ \leff - \liv $,
as a power series in  $ 1 / \Lambda $ . Then
the index
of a graph is equal or smaller than the power of $ 1/ \Lambda $ of the
effective operators present in the diagram. For example, graphs with
a single insertion of a dimension six operator will have indices $ \le 2 $
which implies that the corresponding divergences (if any) will renormalize
operators of index $ s_n  \le 2 $, \ie, of dimension $ \le 6 $.

Other choices for $u$, such as
$ u = 1 $ (for which the ordering is independent of the
number of fermionic legs) will not
be examined further. I will concentrate in the two cases
$ u = 4 $ and $ u = 2 $.

This discussion is necessary when effective lagrangians are
to be used in perturbative calculations since an ordering
consistent with the loop expansion is then required. It is
possible to extend these considerations to
other non-perturbative expansions, such as the large $N$
approach (for a review see Ref.~\colemanN). For simplicity
these possibilities will not be considered here.

\section{{\bfit{The \dec.}}}

In this subsection I will study the \dec\ when
$ \liv $ equals the \sm\ lagrangian, and the
symmetries in $\leff $ are those of the \sm\ (both local
and global). The set of operators
of dimension $ \le 6 $ constructed using the \sm\ fields was given
long ago in Refs.~\burges, \keung, \bw.
Given the (assumed) absence of
right-handed neutrinos, it is easy
to show that there are no operators of dimension five due to the
various (global) symmetries. There
are 81 independent operators of dimension six,~\footnote{f}{For one family
of fermions.} in the sense that
all other operators of dimension six are either a linear
combination of them, or differ by terms which do not
affect the S-matrix (see section 5 for a
discussion of this last point). The whole collection can be found
in Refs.~\bw; in appendix A the operators relevant for
the three gauge boson couplings are discussed. As a simple example
I present here those operators which modify the $S$ and $T$ oblique
parameters~\refmark{\peskin} \con $$ \eqalign{
S&: \quad \ocal_{ W B } = \phi^\dagger \sigma_I \phi \; W_{ \mu \nu}^I
B^{ \mu \nu } ; \cr
T&: \quad \ocal_\phi \up 3 = \left| \phi^\dagger D_\mu \phi \right|^2 .
\cr } \eqn\decst $$

\section{{\bfit{The \ndec.}}}

When $ \liv $ is non-renormalizable the situation which I will consider,
as mentioned previously, corresponds to the case where the terms in
$ \leff $ are ordered according to their value of
$ d_n + f_n/2 - 2 $ (where $ d_n $ is the number of
derivatives and $ f_n $ the number of fermion fields), corresponding to
the choice $ u = 2 $ in \indeq. For the purely bosonic
sector this corresponds to a derivative
expansion and suggests
that this is a relevant ordering when the scalar sector comprises
only the Goldstone bosons that generate the masses for the gauge bosons
(recall that the Goldstone bosons couple derivatively). This is further
supported by the fact that in this case the scalar sector
contains vertices with an arbitrarily large
number of Goldstone fields (see below).

Of the possible applications of this ordering, there are, as
already mentioned, two cases of phenomenological interest.
In the first the low energy particle content is the same
as that for the \sm\ with the exception of the Higgs particle, which
is assumed to be absent or heavy.~\refmark{\chiral}
In the second case the top
quark is assumed to be heavy and the \sm\ particle content, including
the Higgs, generates the low energy spectrum.~\refmark{\yao} Of course these
possibilities can be mixed into the heavy top + heavy Higgs scenario.
The heavy top case was briefly described above, I refer the reader to
Refs.~\yao, for a full discussion. I will concentrate on the
heavy Higgs scenario for the rest of this section.

To motivate the form of the heavy Higgs
Lagrangian recall that one can describe
the scalar sector in the \sm\ using the
matrix $$ \Omega = ( \phi , \tilde \phi ) ,
\eqn\omdef$$
where $ \tilde \phi = i \sigma_2 \phi^* $. It is always possible to write
$$ \Omega = {
 H + v  \over \sqrt{ 2 } } \; U,  \eqn\eq $$ where $H$ is the Higgs field,
$v$ its \vev,
and $U$ is a unitary matrix constructed from the Goldstone boson fields
$ \pi^i$, $$ U = \exp \left( i { \sigbf \cdot
\pibf \over  2 f } \right ) , \eqn\eq $$ where $f$ is a constant (the ``decay''
constant) with units of mass. Using the wave function
renormalization freedom of the $ \pi^i $ I can take $ f = v $.
(see for example Ref.~\itzykson).
$U$ transforms under $ \su2_L \times \ui_Y $ as $$
U \rightarrow V_L U \; e^{ i \sigma_3 \alpha /2 } \eqn\eq $$
where $ V_L \in \su2_L $ and $ \alpha $ is the parameter of the $ \ui_Y$
transformation.

If the Higgs particle is removed then, up to a multiplicative constant,
$ \Omega \rightarrow U$. One can now forget the above derivation
and {\it postulate} that the low energy particle spectrum comprises the
\sm\ fields with $U$ replacing the scalar doublet.
This is the most
economical way of describing the symmetry breaking sector without any
remaining physical scalar excitations.~\footnote{g}{This
choice is made for simplicity only; there
is no loss of generality since all other
possibilities are equivalent.~\refmark{\coleman}}
The penalty for this
simplicity is the non-renormalizability of the resulting
theory which implies that the low energy effective
lagrangian must contain all operators involving
$U$, the \sm\
fermion and gauge fields, and  their derivatives.~\refmark{\chiral}
The fact that the vertices of $ \leff $ involve $U$, which
contains arbitrary powers of the Goldstone fields, implies
that the ordering in \leffordering\ should be such that the index
is independent of the number of boson legs, hence the requirement
$ u = 2 $ in \indeq.

The operators of index zero without fermion fields are
$$ \tr \left\{ \left( D_\mu U \right)^\dagger \left( D^\mu U \right)
\right\} , \qquad
\tr  \left\{ \sigma_3
\left( U^\dagger D_\mu U \right)^2 \right\} , \eqn\chiralnofer $$
whose significance
is most obvious in the unitary gauge: if we set $ U = 1 $ then the
first operator gives the mass terms for the gauge bosons.
The second
operator gives a term violating the custodial
symmetry~\refmark{\sikivie} and generates a modification of the oblique $ T $
parameter, see appendix B.

The operators of index zero and two fermion fields are the fermion
kinetic energy terms
$$ \bar \psi \not \!\! D \psi. \eqn\chiraltwoferi $$
There is also a series of operators containing two fermions together
with $ D_\mu U $,~\refmark{\peccei} these are given in appendix B.
Note that there are terms
which apparently have index $ s_n
= - 1 $: the fermion mass terms in $ \leff $,
$$ m \bar \psi_L U \psi_R , \eqn\chiraltwoferii
$$ which would ruin the consistency
between the ordering in $ \leff $ and the loop expansion.
These terms appear always together with
the covariant derivative terms  \chiraltwoferi\ and so it is consistent to
assume that the mass parameter $m$ has index one,
which I will henceforth adopt. This is expected
to be qualitatively correct
as long as $m$ can be kept naturally light (imposing, for example,
a chiral symmetry).
Then \chiraltwoferii\
will have index zero. This assumption is further discussed in Ref.~\gasser.

The operators of index zero with four fermion fields and no $U$ fields
can be found in Ref.~\bw. The remaining possibility corresponds
to the
four-fermion operators containing one or more $U$ fields. These
can be constructed from the previous four-fermion operators
by judicious insertions of the object $ U \sigma_3 U^\dagger$.

There are 19 operators of index two with no
fermions~\refmark{\appelquistwu} given in appendix B. These have canonical
dimension $ = 4 $
and their coefficients are $ \Lambda $ independent (up to
logarithmic corrections). At tree level their effects do not
vanish as $ \Lambda \rightarrow \infty $ despite their being
generated at this scale.

\section{{\bfit{Comments.}}}

It is worth emphasizing that
the expansion of $ \leff $ according to the index of the operators
is very different depending on the choice of $u$. For example, in the
\dec\ the operator $
W_{ \mu \nu }^I W_{ \nu \rho}^J W_{ \rho \mu }^K \epsilon_{ I J K } $
\con\ has index $ s_n = 2 $ and in included in the catalogue of Ref.~\bw.
In contrast, for the \ndec, this operator, having six derivatives
(recall that each field tensor is interpreted as a derivative commutator
and so counts as two derivatives)
and no fermion fields, has index four and is not included in the list
of appendix B, being a higher order correction than those operators
generated by $ \liv $ at one loop.

A natural question to ask regarding the range of applicability of
the effective lagrangians described above is how close to $ \Lambda $
can we trust the parametrization in terms of effective local operators.

For the \dec\ this depends on the characteristics of the underlying
physics, namely, whether the heavy excitations have wide or narrow
resonances. If wide, these resonances will have long
``tails'' and their effects will intrude into what one would naively
consider the region where the effective lagrangian parametrization should
be valid. This generates anomalously large coefficients
(up to an order of magnitude in some cases) for some of the operators,
compared to the values expected from naturality (see section 4.2).

In the \ndec\ the above arguments indicate that, for those terms without
fermions, $ \leff $ is organized as a (covariant) derivative expansion.
We will see in the next section that the coefficients of an $n$ derivative
operator is proportional to $ 1/ ( 4 \pi v )^n  $. In the case
of the Higgs-less \sm\ $ v \simeq 246 \gev $ so that the
effective lagrangian parametrization will certainly break down at
energies of the order $ 4 \pi v \sim 3 \tev $ in this case.

The question now arises as to the significance of this scale.
It was argued in Ref.~\dougangolden\ that, since $ \leff $ will
generate all required physical amplitudes with the correct unitarity cuts
as long as the energies are below $ 4 \pi v $, the breakdown of this
expansion must be associated with a new mass threshold at
or below this scale. In this argument strong use is made of the
naturalness requirement~\refmark{\weinberg} (see section
4.1) that $ \leff $ is an expansion in $ \partial / ( 4 \pi v ) $.
Reference~\dougangolden\ also provides various examples in which
these arguments are illustrated. The arguments can also be
applied to low energy pion physics, where the scale
$ 4 \pi v $ is remarkably close to the mass of the $ \rho $
meson.~\refmark{\chivukula}

These statements have been criticized in Ref.~\appelquistterning\
on the basis that the breakdown of the expansion in a derivative
series only implies that one must sum terms of all orders before
deriving any conclusions, and that there are functions (which may
represent Green's functions) for which the derivative expansion
has large coefficients but still may be summed into an analytic
result. The derivative expansion in the example provided, however,
does not have the same order of magnitude coefficients as required
by the  arguments in Refs.~\dougangolden, \chivukula, but are
instead much smaller.

\chapter{Orders of Magnitude.}

In this section I will discuss various issues surrounding the magnitude of
the effects produced by operators of dimension greater than four. I will
first consider the expected order of magnitude for the coefficients of the
effective operators in the \ndec. Then I will turn to the same question
in the \dec.

\section{{\bfit{The \ndec.}}}

In order to determine the order of magnitude of the coefficients
of the operators
presented, for example, in appendix B, it is natural to
require that the radiative corrections
to the coefficient of an operator be at most as large as its tree level
value. Equivalently one can argue as follows:~\refmark{\weinberg,\georgibook}
when considering the renormalization of the operators in $ \leff $
it is found that all the coefficients are in general renormalized.
It is then understood that the renormalization group running of
these quantities can be used to evolve them in energy,
up to the scale where
a mass threshold is crossed (presumably at a scale
$ \sim \Lambda $).~\refmark{\georgi} At this point the effective coefficients
are determined by the couplings of the heavy theory.
Then it is natural to require that,
if the renormalization scale is changed
by a factor of order one, the running parameters also change by a factor
of order one. This requirement (in either of its forms) determines
the order of magnitude for the coefficients; for a discussion on the
assumption of naturality see section 7.3.

The procedure which I will follow to determine the coefficients of $ \leff $
appeared in Ref.~\georgimanohar. I will
assume that the theory is explicitly gauge invariant, either because
it was originally so, or because it was rendered gauge invariant using the
procedure outlined in section 2.1.

Consider a theory with scalar fields $ \phi $,
fermionic fields $ \psi $  and gauge bosons $W$. Then the
relevant vertices have the symbolic form
$$ \Lambda^4 \lambda ( \phi / \lb )^A (\psi / \lf^{3/2}
)^B (p / \Lambda )^C ( g W / \Lambda )^D ,
\eqn\strongvertex $$
where $p$ represents a derivative, $  \Lambda $ is a UV cutoff,
$ \lambda $ is a coupling constant, and the other scales,
$ \Lambda_{ \phi , \psi } $, are to be
determined. Since $ \Lambda $ is associated with the momentum
scale I divide
$p$ (a generic momentum) by this scale; since gauge fields appear only in
covariant derivatives, they are divided by the same scale. The
quantities $A, \ B,  \ C $ and $D$ are assumed to be integers.
Since the $W$ fields appear always inside a covariant derivative
it is sufficient to consider vertices with $ D =  0 $
(field tensors are treated as covariant derivative commutators).

I will assume that the gauge-boson couplings to
fermions and bosons (including
self couplings) are $ \lesim 1 $; based on this assumption
I neglect internal gauge
boson lines, as well as Fadeev-Popov ghost lines (this
refers to the \sm\ gauge bosons, the ones
of the underlying theory --- if any --- are already
``integrated out'').
Note that since we are dealing with a gauge theory,
I can choose a gauge in which the
$W$ propagator which drops off at large momentum.

Now consider a graph with $V$ vertices which generates an
$L$ loop correction to \strongvertex. This contribution
will be of
the same order provided (I replace all loop momenta by
$ \Lambda $ since we are interested only in an order
of magnitude estimate)  $$ 1 \sim (
\Lambda^4 \lambda )^{ V -1 } \lb^{ A - \sum A_i } \lf^{3( B- \sum B_i )/2 }
\Lambda^{ C - \sum C_i }
\Lambda^{ \sum C_i - C + 4 L - 2 \ibos -\ifer }
( 4 \pi )^{ - 2 L } , \eqn\inter $$ where
$i$ labels the vertices in the graph and $ \ifer $ ($ \ibos $)
is the number of internal fermion (boson)
propagators. Using the relations $ \sum A_i = A + 2
\ibos $ and  $ \sum B_i = B + 2 \ifer $ \inter\ becomes
$$ ( 16 \pi^2 \lambda )^{ - L }
\left( { \lambda \Lambda^2 \over \lb^2 } \right)^\ibos \left( { \lambda
\Lambda^3 \over \lb^3 } \right)^\ifer \sim 1 ; \eqn\eq $$
this requires $$ \lambda \sim
\inv{ 16 \pi^2 } ; \qquad \lb \sim \inv{ 4 \pi } \Lambda ;
\qquad \lf \sim \inv
{ ( 4 \pi )^{ 2/3 } } \Lambda . \eqn\eq $$
Substituting back into \strongvertex, and using the fact that gauge bosons
appear always in a covariant derivative denoted by $ \dcal $, yields
$$ {\Lambda^4 \over ( 4 \pi )^{ 2 - A - B } }
\left( { \phi \over \Lambda } \right)^A \;
\left( { \psi \over \Lambda^{ 3/2 } } \right)^B \;
\left( {\dcal \over \Lambda } \right)^C . \eqn\strongvertexi $$

If the scalars appear in the form of a
unitary field,  $$ U \sim \exp ( i \phi / \lb )
\sim \exp ( 4 \pi i \phi / \Lambda
) , \eqn\eq $$ then the vertex takes the
form $$ \inv{ ( 4 \pi ) ^{ 2 - B } }
\Lambda^4 { \dcal^C U^{ A'} \psi^B \over
\Lambda^{ C + 3 B / 2 } } , \eqn\chirvertex $$ where
$A'$ is an integer and, as above, $
\dcal $ denotes the covariant derivative. Note that,
as claimed in section 3.2, this implies that the bosonic sector
of $ \leff $ is organized as an expansion in powers of $ \dcal / \Lambda $.

The operators in \chiralnofer\ have coefficients of order $
( \Lambda / 4 \pi )^2 $. The first of these operators,
$ \tr | D_\mu U |^2 $
corresponds to $ B = 0 $, $ A' = C = 2 $, and
generates the masses of the gauge bosons (this is obvious in the
unitary gauge). It then follows that
$$ \Lambda \simeq 4 \pi v \sim 3~\tev \qquad \hbox{(\ndec)};
\eqn\chirlimit $$
where $ v \simeq 246 \gev $ corresponds to the scalar \vev\ in the
\sm. The other operator in \chiralnofer\ should also have a coefficient
of order $ v^2 $ according to the previous argument; in fact,
experimentally it is found that its
coefficient is very much suppressed, being proportional to the
deviation of the $ \rho $ parameter from one. This illustrates a
caveat in the above estimates: the results \chirvertex\ corresponds
to the largest radiative corrections allowed by naturality arguments.
But in specific situations the coefficients
can be suppressed due to some unknown effects in
the underlying theory. Such a situation is envisaged in Ref.~\gounarisi\ where
the heavy physics is assumed to respect the $ \su2_R$ symmetry
of the scalar sector in the \sm, with this assumption all operators of
dimension$ \ge 6 $ violating this symmetry can be ignored.

The operators in \chiraltwoferi\  have coefficients
of order one. Then, by the discussion below
this equation, the same will be the case for the
operators in \chiraltwoferii. Note that it is possible to impose a
chiral symmetry to protect the fermion masses; in this
case one can naturally assume that they are $ \ll
\Lambda $.~\refmark{\thooft}

Similarly the four-fermion operators have coefficients of order
$ ( 4 \pi / \Lambda )^2 \simeq 1 / v^2 $. This is
different from the usual assumptions~\refmark{\eichten, \ruckl, \ehlq,
\djouadi,\doncheski,\barbiellini} where the coefficient is taken to be $ 4 \pi
/ \Lambda ^2 $.
This is argued on the basis that the coefficient is of the form
$ g_H ^ 2/ \Lambda ^2 $ for some coupling $ g_H $; assuming the
underlying theory is strongly interacting implies $ g_H^2 $ is large,
and this is presumed to mean that $ g_H^2 = 4 \pi $, \ie\
$ \alpha_H = 1 $, which
taken as the {\it definition} of a strongly
interacting theory. This is of course a fuzzy statement: one could
equally define a strongly interacting theory by the condition
$ \alpha_H = 4 \pi $, which is further supported by the arguments above.
Note however that the assumption $ \alpha_H = 1 $
does not necessarily contradict the
above naturality value for the coefficients since,
as mentioned above, they can be suppressed.~\footnote{h}{A more serious
problem arises when the kinetic energy for the gauge bosons
is considered. This corresponds to $ A ' = B = 0 $ and $ C = 4 $ in
\chirvertex\ which implies a coefficient $ \sim 1 / 16 \pi^2 $ instead
of the correct $ 1 / 4 $; see Ref.~\georgimanohar\ for a discussion on
this point.} Finally note that all index $ s_n = 2 $ operators in
appendix B have coefficients of order $ 1 / 16 \pi^2 $.

These estimates can be carried over to the one example of a strongly
coupled theory to which this approach has been applied, and for which
there are abundant experimental results: the low
energy strong interactions. The results (see Ref.~\pich\ for a recent review)
are perfectly consistent with the above estimates (for example,
the {\it measured} values for the coefficients of the four derivative
operators are all of order $ 1 / 16 \pi^2 $).

\section{{\bfit{The \dec: tree vs. loop-generated operators.}}}

\subsection{Strongly coupled case}

In the \dec\ and when the underlying physics is strongly interacting,
one might attempt to follow the same arguments as in the previous section
and assert that \strongvertexi\ should remain valid. There is
a serious problem with this reasoning: if we consider the mass
terms for the scalars, corresponding to $ A = 2 $, $ B = C = 0 $ in
\strongvertex, it follows immediately that the  masses are
$ O ( \Lambda ) $.~\footnote{i}{The same appears to be true
for the fermion masses.
This disaster can be avoided imposing a chiral symmetry which
insures that, for the case $ B = 2 $ , $ A = C = 0 $, \strongvertex\
will acquire a factor $ m / \Lambda $, where $m$ is the tree level
fermion mass.}
The scalar fields should, in this case,
not be included in the low energy spectrum,~\footnote{j}{This problem
does not
arise for the \ndec\ due to the Goldstone nature of the scalar fields
in that scenario.} and we are lead back to the \ndec.

It is possible to devise models where there are cancellations
among diagrams thus allowing $ m \lowti{ scalar} \ll \Lambda $;
an example is given by imposing supersymmetry on the model.~\refmark{\gates}
Such theories invariably generate light particles not present
in the \sm\ and will be probed directly
in the next generation of colliders; their study using an effective
lagrangian approach requires then the modification of the low
energy spectrum. In order to keep the discussion at a manageable level I will
ignore this possibility in the following.
Thus I will assume
in the remainder of this paper that the \dec\ is associated with
a weakly interacting underlying theory, which I discuss next.

\subsection{Weakly coupled case}

For the \dec, and when the underlying theory is weakly coupled
the considerations used
previously to determine the operator coefficients are not applicable
due to the assumption that the couplings are small: higher loops
only give small corrections. The relevant
question now becomes, which operators
can be generated at tree level by the underlying theory? This is so because
loop-generated operators will acquire a loop suppression factor of $ \sim
1 / 16 \pi ^2 $ which is absent for the tree-level operators
(this will be discussed further below), hence their effects
are quantitatively much smaller. It must also be kept in mind
that the operator coefficients will also contain
small ($ \lesim 1 $) coupling constants.

In the discussion below
I will consider the interesting case where the low energy physics is
described by the \sm\ and assume
that the heavy physics is described by a gauge theory.
Before proceeding I will standardize notation. In the full theory
I separate the gauge indices corresponding to the low energy gauge (\sm)
group, denoted by $ a , b $, etc., from
the remaining gauge indices,
denoted by $A , B $, etc. The structure constants of the group in the full
theory are denoted by $f$. Light gauge bosons are denoted by $W$,
heavy ones by $ \wcal $;
light scalars are denoted by $ \phi $,
heavy scalars by $ \Phi $; similarly light fermions are denoted by
$ \psi $ while their heavy counterparts by $ \Psi $.

Formally what is being done is ``to integrate
out'', to lowest order in $ \hbar $,
all the heavy fields. The gauge fixing in the heavy theory is
such that the resulting effective action is manifestly gauge
invariant.~\refmark{\einhornandwudka}
Thus, for this calculation, all light fields are
external and hence I can assume that the light
indices are unbroken (the light gauge
group is the unbroken subgroup). The broken generators
(those with indices $A,B,$ etc.) carry a representation of the unbroken
group (see for example Ref.~\wudka). In the calculations below I will
assume for simplicity that the heavy physics has no
super-renormalizable vertices, and I will use the fact that there are
no dimension five operators that can be constructed from \sm\
fields satisfying the \sm\ symmetries; for a full discussion of the
general case see Ref.~\arztii.

The gauge structure of the full theory disallows certain vertices.
Consider first  a vertex with
two light gauge bosons and one heavy one, it is proportional to
$ f_{ a b C } $ which vanishes since the unbroken generators form
a Lie algebra: the $ W^2 \wcal $ vertices are absent.
Similarly there are no $ W^3 \wcal $ vertices. Vertices of the type
$ W^2 \Phi $ or $ W^2 \phi $ are also not present since they contain
two unbroken generators and they must be proportional to a \vev.
Vertices of the type $ W^2 \phi \Phi $ and
$ W \psi \Psi $ are also not allowed. Finally vertices of the type
$ W \wcal \Phi $ are absent since the fields $ \Phi $ are orthogonal to
the Goldstone boson directions.

With these restrictions it is easy to see that there are no tree level
graphs with only $W$ external legs corresponding to operators of dimension
six. For graphs with some $ \psi $ or $ \phi $ external legs I
use the fact that the final result will be gauge invariant in order
to consider only those graphs with no external $W$ legs: these will be
generated by replacing derivatives with covariant derivatives.
It is then a matter of patience to prove that the
only tree-level operators of dimension six (for a
theory with super-renormalizable vertices the same results are
obtained~\refmark{\arztii}) are
$$ \matrix{
\left( \phi D \phi \right)^2 ,
&  ( \phi \bar \psi ) \not \!\! D ( \phi \psi ) ,  & \phi^3 \bar \psi \psi
\cr
\left( \bar \psi \psi \right)^2 ,
&  \phi^6 ; &  \cr } \eqn\otree $$ (understood to be a schematic
description of the relevant operators).

The operators of the type \otree\ will be suppressed only by two
coupling constants: in  $ ( \bar \psi \psi )^2 $
the coupling for the vertex $ \bar \psi \psi \Phi $ occurs twice; in the
remaining operators containing fermions the
coupling for the vertex $ \bar \psi \phi \Psi $ occurs twice; fin
$ \phi^6 $ the coupling of the vertex $ \phi^3 \Phi $
appear quadratically, etc.
Among the operators in Ref.~\bw\ all but the
above will have coefficients $ \sim 1 / 16\pi^2 $ (see section 4.4).

\subsection{Higher dimensional operators}

Heretofore I have concentrated the discussion on dimension six operators.
There are, however,
situations where higher dimensional operators are non-negligible
and can even be dominant.~\refmark{\arztii}
This occurs if, for the \dec, there are
contributions to a certain observable from both
loop-generated dimension-six operators, and tree-generated dimension-eight
operators. The suppression in the first case is $ \sim 1 / 16 \pi^2 $,
while in the second it is $ \sim v^2 / \Lambda^2 $, where $v$ is the
\sm\ scalar \vev. It follows that for scales below $ 4 \pi v
\sim 3 \tev $ one must
include all contributions from tree level
generated dimension eight operators in any calculation for which the
contributing dimension six operators are loop generated.

\section{{\bfit{Predictability in effective theories.}}}

The often-claimed defect of theories with terms of mass dimension
greater than four lies, not in their lack of renormalizability
(since they {\it are} renormalizable, see section 7),
but in the presumed lack of predictability. This is in fact not the case.

To see this recall that, as
was emphasized in section 3, it is possible to order
the terms in   $ \leff $ according to their index, this ordering
being consistent with the loop expansion.
Suppose now a certain set of observables is calculated using only
operators whose indices are below a certain upper value. It is
then clear that we are dealing with a theory with a finite number of
parameters and so with non-trivial content. Moreover one can use
\strongvertex\ or \chirvertex\ to estimate
the corrections generated by terms of higher indices;~\refmark{\polchinsky}
the corrections are, within the range of applicability of the
effective lagrangian parameterization, subdominant.
The  number of terms
in the expansion that must be kept depends on the level of precision the
experiment has achieved.
Similar arguments can be used in the \dec.

Thus, despite having an infinite set of parameters, effective
theories do have content (a fact routinely used
in the chiral lagrangian approach to the strong interactions
at low energies~\refmark{\gl, \pich}). Effective lagrangians
are provisional models whose
parameters we expect to be able to deduce from a small set
of constants appearing in a more fundamental theory; but, as
long as we have no direct evidence of the new interactions, this
way of parametrizing is very appealing as it is theoretically
consistent and model and process independent.

\section{{\bfit{Naturality in effective theories.}}}

It could be argued that naturality is in itself an extra assumption
which has no scientific origin, but an \ae sthetic one. If one disposes of
naturality on the basis of being as open minded as possible, then one
can include any (Lorentz invariant) term in $ \leff $ with completely
arbitrary couplings.

This type of models is useless since it completely lacks
predictive power.
Since the model has absolutely no constraints on the coefficients,
no relationships are obtained between the various terms in the
lagrangian; the fact that experiments confirm various connections between
such coefficients must be assumed to be coincidences within this
approach. For example the fact that $ \rho = 1 $ at tree level in the
\sm\ would be an interesting fact, but of no deep significance since the
effective lagrangian would have terms of the form $ m_W^2 W^+ \cdot W^- $
and $ m_Z^2 Z^2 /2 $ with no connection between $ m_W $, $ m_Z $ and the
weak mixing angle derived, for example, from the neutrino cross sections;
even if this connection is imposed at tree level, it will be spoilt,
and very strongly, by loop corrections. Similar arguments could be
constructed, for example, for lepton universality.

The point is that one cannot have it
both ways: if the \sm\ is assumed as an excellent first
approximation, then the corrections should be such that this assumption
remains valid. This requires the presence of gauge invariance and
a hierarchy on the coefficients of the effective operators. If
either of these conditions is violated the consistent evaluation of
all low energy processes will generate a theory different from the
one originally considered. For example, most or all
of the masses would receive corrections of $ O ( \Lambda ) $
(sections 4.5 and 7.3), and the couplings become large (section 7.3).

\section{{\bfit{Loop factors.}}}

In this sub-section I re-examine the claims made above that
the coefficients of loop-generated
operators are subdominant. This discussion will be limited to
the case where the heavy physics is weakly coupled.

It is clear that, generically, any  single loop contribution is accompanied
by a factor of order $ 1 / 16 \pi^2 $.
One must also remember that
the heavy loops will contain not only these loop factors but also
a certain number of small coupling constants. Still, it is often argued that
this suppression can be reduced when many graphs contribute to the same
physical quantity, provided the
contributions add coherently. One could envisage, for
example, integrating out $N$ identical heavy fermions with
$ N \gesim 160 $, in which case the total loop
contributions would be unsuppressed. The simplest view one can take
for this type of situation is that, since tree-level and one-loop
contributions are comparable, the loop expansion is not useful
in analyzing this type of model, which are not manageable using
perturbative techniques (note that it is not assumed that
the couplings are suppressed by factors of $ 1/ N $ and so the large
$N$ expansion will also not be useful).

Of particular relevance are the contributions from this large number of
particles to the vacuum polarization  of the light excitations.
{}From the fact that the gauge boson masses are protected \`a la
Veltman,\refmark{\veltmaniruv}
it follows that the presence of $ \sim 16 \pi^2 $ virtual loops
adding coherently
will give a large correction, of order 100\%\ to the gauge boson masses
(see section 7.3 for a related discussion).
The scalar masses, in contrast, are not necessarily protected and
a correction of order $ \Lambda $ is expected (see the example
below for a specific model). It follows
that under these circumstances
all scalars coupling to a large number of heavy fields
would become heavy through radiative corrections, radically modifying
the low energy theory. An exactly
solvable example of this situation is presented next.

\subsection{A simple example.}

To see whether these somewhat qualitative arguments are quantitatively
correct, it is convenient to study an exactly solvable model.
I chose a theory of $N$ degenerate fermions interacting with a
scalar field $ \theta $ in two space-time dimensions.

The lagrangian is
$$ \lcal  = \half ( \partial \theta )^2
+ i \; \sum_{ a = 1 } ^N \bar \psi_a \dcal \psi_a ;
\qquad \dcal =
\not \! \partial + { i g \over 2 } \gamma^\mu ( \partial_\mu \theta )
- i m e^{ i g \theta \gamma_5 }\eqn\eq $$ where $g$ is a coupling constant
and $m$ a heavy mass.

I will be interested in the effective theory obtained by
integrating out the fermions.  A chiral rotation
in the fermions changes $ \theta $ and induces a Jacobian, which is
evaluated using Fujikawa's method.~\refmark{\fujikawa}
The result, described in appendix C, is
$$ \leff  = \half ( \partial \chi )^2
+ { N m^2 \over 4 \pi } \cos ( \lambda \chi );
\qquad \chi = { 2 g \theta + \pi \over \lambda } , \quad
\lambda = { 2 g \over \sqrt{ 1 + g^2 N / ( 4 \pi ) } } \eqn\eq $$

This is the Sine-Gordon model extensively studied in the
literature (for a review  see Ref.~\rajaraman).
The mass of the linear excitations
is $ 2 m / c $ where $ c = \sqrt{ 1 + 4 \pi / ( g^2 N ) } $,
the solitons present in this model have mass
$ m N c / \pi $,\refmark{\rajaraman}
etc. Since $ c \ge 1 $ the solitons are always heavy;
light scalar states appear only when $ c \gg 1 $, which requires
$ g^2 N \ll 4 \pi $.
In the limit where $N$ offsets the loop
suppression factor $ g^2 / 4 \pi $, the quantity $c$ is of order one
and there are no light excitations at all.

These support the previous claims that whenever the number of graphs
is so large as to compensate the loop suppression factor (including
the small coupling constants) a strongly coupled theory results; in
this case the loop expansion is not reliable.
The presence of a large number of particles
in the loops can indeed overcome the loop suppression factors, but in this
case the low energy theory is completely different from the one
naively expected.

\chapter{Equations of Motion and Blind Directions.}

\section{Equations of motion}

The classical equations of motion can be used to
reduce the number of operators in $ \leff $.~\refmark{\bw,
\derujula,\arztiii} This is based on the
fact that if two operators $ \ocal $ and $ \ocal ' $ are such that
$ \ocal - \ocal' = \acal (\delta S / \delta \chi ) $, where
$ \acal $ is some local operator, $S$ is the classical action and $ \chi $
a field of the theory, then an insertion of $ \ocal $ gives the
same S-matrix element as that of $ \ocal' $ (though the Green's
functions {\it are} different). A simple proof of this fact is
presented in appendix D; for a thorough discussion see Ref.~\arztiii.
Such operators will be called {\it equivalent}.

Thus, if $ \leff $ has a term $ \eta \ocal + \eta' \ocal' $
where $ \ocal $ and $ \ocal' $ are equivalent,
then, to lowest order in $ \eta $ and $ \eta' $,
all physical quantities will depend on $
\eta + \eta' $ but never on either of them independently.
A term in $ \leff $ is {\it redundant} if it is equivalent to another
term already in the effective lagrangian. In Ref.~\bw\ and many
other publications great efforts are made to eliminate all
redundant operators.

There is an important point which is obscured by eliminating
redundant operators. Consider for example the case where the
heavy physics is weakly coupled, and suppose $ \ocal $ and
$ \ocal' $ are equivalent. It is then possible for the heavy physics to
generate, for example, $ \ocal $ at tree level, and $ \ocal' $
at one loop. Thus if we eliminate $ \ocal $ in favor of $ \ocal' $
and estimate that coefficient for $ \ocal' $ has its natural size,
we would severely underestimate the heavy
physics effects : there is in fact a very significant contribution from $ \ocal
$
unencumbered by loop suppression factors.

As a concrete example consider the operator
$$ \left\{ ( D_\mu \phi)^\dagger ( D_\nu \phi ) + \half \phi^\dagger [
D_\mu , D_\nu ] \phi \right\} B^{ \nu \mu } \eqn\eq $$ which,
in the \dec\ for a weakly interacting heavy theory,
is generated
only via loops (see section 4.2).
On the other hand the use of the equations of motion show
that it is equivalent to $ ( \phi^\dagger D^\nu \phi)
j_\nu \up B $ where $j_\nu \up B $ is
the source current for $B$;
this last operator can be generated at {\it tree}
level. Therefore, even if the S-matrix elements cannot
distinguish between the
first and second operators, there is a very large quantitative difference
whether the underlying physics generates the second one or not.

In view of the above it appears to be more useful not to eliminate
redundant operators from $ \leff $; then one can
always estimate the contributions from any given operator without having to
keep track of the expected order of magnitude of the contributions
from other operators equivalent to it. One may also exploit the redundancy
of operators to check calculations of S-matrix elements.

\section{Blind directions.}

Another consideration relevant for quantitative estimates is the
possible presence of ``blind directions'':~\refmark{\derujula}\
these are operators
to which we have no experimental sensitivity since they affect
quantities only at the one loop
level (or beyond). An example is the operator
$$ \ocal_W = \epsilon_{ I J K  } W_{ \mu \nu }^I W_{\nu \lambda }^J
W_{ \lambda \mu }^K . \eqn\ow $$

In some publications (see, for example, Ref.~\derujula) it has been
claimed that there is no fundamental difference between the blind
operators and the ``sighted'' ones. One can then translate any limits
on $ \Lambda $ obtained by using a sighted operator into limits
on the effects of a blind operator. If correct, this argument
implies that the limits derived from the
measurements at LEP1 are so severe that LEP2 will be almost
insensitive to any kind of new physics. I will now show that
this claim is in fact an extra assumption strongly
dependent on the heavy physics. Counter-examples are readily available.

Consider first the following toy model
consisting of a light
scalar field $ \phi $ interacting with two heavy fermions $ \psi_a \ (a = 1 ,2
)$. The lagrangian is $$ \lcal = \underbrace{ \half ( \partial \phi )^2 - \half
m^2 \phi^2 - \inv6 \sigma \phi^3 - \inv{24} \lambda \phi^4 }_{\hbox{{\sanser
light \ sector }}} + \underbrace{ \sum_{ a = 1 }^2 \bar \psi_a
\left( i \not \!
\partial - M + ( - )^a g \phi \right ) \psi }_ {\hbox{{\sanser heavy \ sector
}}} \eqn\eq $$ When the fermions are integrated out,
they produce an effective
action that is even in $ \phi $.
If low energy ($ \ll M $)
experiments are sensitive to, for example, $ \phi^5 $ but
not to $ \phi^6 $ (which is then a blind direction),
there would be no experimental indication of the heavy sector.
The null results could be interpreted as
$ \Lambda $ being astronomical if the above assumptions are made,
while in fact a new generation of
experiments could very well uncover the presence of the heavy fermions.

This would have little more than academic interest
if the same argument
could not be applied to realistic models. In fact this is not the case.
Suppose we add to the \sm\ two vector-like
fermion doublets $ \Psi_a , \ a=1,2 $,
which have a common mass $M \gg v$.
Suppose one doublet
has hypercharge $y$ the other $ - y $. If we integrate these fermions
out we generate a series of one loop graphs with external $B$ and $W$ legs
\con. The choice of masses and hypercharges guarantees that all
graphs with an odd number of $B$ legs will cancel out.~\footnote{k}{If the
fermion masses are split by an amount $ \Delta M \ll M $,
the terms with an odd number of $B$ legs are suppressed by
a power of $ \Delta M / M $.} This implies that,
for example, there will be no contribution to the oblique
$S$ parameter,~\refmark{\golden} see \decst. If we
then take the point of view advocated in Ref.~\derujula,
we would conclude that the scale of new physics is enormous, and
that there is no hope in the near future of being sensitive to,
for example, the physics that generated the operator \ow.

There is no mystery in the above results: unknown symmetry properties
of the heavy physics may forbid certain operators. If we have, accidentally,
a preference towards measuring the effects of precisely the
suppressed operators,
we would then (erroneously) conclude that the scale of new physics is
extremely high. This could be rephrased by stating that
the scale responsible for a certain set of operators could
be very high, but we
cannot extend this to the whole manifold of possible
operators; though this is often assumed, it remains just that, an assumption.
When evaluating the results of a certain
experimental search this fact must be kept in
mind: though factors such as the loop suppression factors do restrict
our sensitivity to new physics (see section 8),
it is conceivable that current measurements
are prejudiced, due to experimental constraints, against the strongest
effects from new interactions which may be revealed in future
experiments.

As a last comment on blind directions it is important to point out
that one can always try to dispose of such operators by changing
to a basis where most operators are ``sighted''. To illustrate
this point I will consider the case where the low energy
theory is the \sm\ without fermions and consider the
blind operator~\refmark{\derujula, \escribano}
$ \ocal = B^{ \mu \nu } \; D_\mu \phi^\dagger D_\nu \phi $
(the notation is the same as above). Use of the
equations of motion shows that this is equivalent to the linear combination
$$ { i g \over 4 } B^{ \mu \nu } W^I_{ \mu \nu }
\phi^\dagger \sigma_I \phi +
i g' \left| \phi^\dagger D_\mu \phi \right|^2 +
{ i g' \over 2 } \left( \phi^\dagger \phi \right)
\left| D_\mu \phi \right|^2 +
{ i g' \over 4 } \left( \phi^\dagger \phi \right) B_{ \mu \nu }^2
\eqn\eq $$ plus a renormalization
of the scalar self coupling constant \con.
The first operator affect the oblique parameter $S$, similarly
the second modifies $T$; the third operator modifies the Fermi constant
while the last one affects the $Z$ mass and the weak mixing angle.
The point is that all these operators affect measured
quantities at tree level. Thus we obtain a better bound on $ \Lambda $ by
choosing the second set of operators
than the original one. This result should be tempered, however, by the
previous comments on the magnitude of the coefficients of
equivalent operators.

\chapter{Unitarity.}

Effective interactions vertices have been used to provide examples where
tree level unitarity is violated, and this deserves comment.
First it must be remembered that violation of tree level unitarity does
not mean that the theory violates unitarity, but that the coupling
constants are so large that perturbative calculations are unreliable.
Thus, violation of tree level unitarity signals the onset of a
strongly interacting regime.

It must also be remembered that the use of effective operators carries
the responsibility of not using this parametrization above a certain
scale $ \Lambda $. The question to ask is therefore: given a certain
reaction whose CM energy is below $ \Lambda $, are there violations
of tree level
unitarity? In order to answer this, the results of
section 4 must be used in order to estimate the magnitude
of the coefficients of the effective operators.

For example, in the \dec, a typical
operator $ \ocal $
of dimension six with $ \le 2 $
scalar fields, produces, in a $ 2 \rightarrow 2 $ process,
amplitudes that grow like $ s $ ($s$ is the usual Mandelstam
variable), which has the potential of violating tree level
unitarity. But to actually determine whether
this possibility is realized
it must be remembered that the operator under consideration
appears as a term $ \alpha \ocal / \Lambda ^2 $ in $ \leff $, so that
its contribution to the scattering amplitude  is of order
$ s \alpha / \Lambda^2 $, where $
\alpha  \lesim 1 $ (see section 4).
It follows that these
terms can generate (tree level) unitarity violations
only for energies $ \gesim \Lambda $,
\ie\ beyond the range of applicability of the effective
lagrangian parametrization (at such energies
the higher dimensional operators dominate over the ones included and
the expansion breaks down).

A specific example of this is the following: consider in
the process $ e^+ e^- \rightarrow W^+ W^- $, the
contributions of the effective lagrangian $$
\lcal \lowti{St. Model} - { i e \over m_W^2 } W^+_{ \lambda \mu}
W^-{}^\mu{}_\nu
\left( \lambda_\gamma F^{ \nu \lambda }
+ \cot \theta_W \lambda_Z Z^{ \nu \lambda } \right) , \eqn\eq $$
obtained from Ref.~\hagiwara; $F$ denotes the electromagnetic field
tensor, $Z^{ \mu \nu } = \partial^\mu Z^\nu - \partial^\nu Z^\mu $ and
similarly for $W^\pm_{ \mu \nu } $;
 $ \lambda_{ \gamma , Z } $ are constants and $ \theta_W $ denotes the
weak mixing angle.
If this is rendered gauge invariant along the lines of section 2
then $ \lambda_{ \gamma , Z } \sim g^2 / 16 \pi^2 $ (see section
4.1).

The amplitude for $ e_L^+ e_R^- \rightarrow W_L^+ W_L^- $
is~\refmark{\hagiwara} $$ \acal = e^2
\sin \Theta { \lambda_Z - \lambda_\gamma
\over 4 m_W^4 } s^2 + O ( s ) , \eqn\eq $$
where $ \Theta $ is the $W$ center
of mass scattering angle. The corresponding
$ \ell = 0 $ partial wave violates unitarity when
$$ \sqrt{s } \gesim \left( { 128 m_W^4 \over e^2
| \lambda_Z - \lambda_\gamma | } \right)^{ 1/4 } \simeq
{ 0.5 \tev \over  | \lambda_Z - \lambda_\gamma |^{ 1/4 } } \eqn\eq $$

If the
$ \lambda $ term in the denominator is assumed to
be of order one, then this corresponds to an energy $ \sim 500 \gev $;
when the $ \lambda $  have their natural sizes, this is
increased to $ \sim 2.1 \tev $, which is of the same order as the
scale where the effective lagrangian parametrization breaks down
(see \chirlimit). It is therefore
           inconsistent to
suppose that the coefficients $ \lambda $ are of order one.

Within the \dec\ the above lagrangian is generated by the
operator \ow. In this case
however $ \lambda_Z = \lambda_\gamma $ so that the potential tree level
unitarity violations are generated by the $ O ( s ) $ terms in
$ \acal $ (if at all) and will be much smaller.

This example illustrates the importance of keeping all small factors.
If the coefficients in $ \leff $ are assumed to be unnaturally large
the conclusion would be that there are very significant effects at
scales well within reach of a near-future collider. If the natural
size of the coefficients is retained the effects
are generally found to be very small.

Another case where the study of
unitarity in an  effective theory becomes interesting
is within the context of the so called ``delayed unitarity''
scenario.~\refmark{\ahn}
The idea is that gauge theories often require cancellations
among diagrams to enforce unitarity. If some of these diagrams
are suppressed with respect to others in some energy range, these
cancellations will not be apparent in the said range; this can give
rise, in some cases, to measurable effects. So, for example, in
an explicit calculation~\refmark{\ahn} of the effects of a
hypothetical fourth generation of heavy fermions
in $ e^+ e^- \rightarrow W^+ W^- $, the cross section
acquires a correction factor $ \sim 1 + s/ ( 4 \pi v )^2 $
where $ v \simeq 246 \gev$.  Since this is a perturbative calculation,
it is reliable as long as perturbation theory is valid, which requires
the mass of the fermions to be below $ \sim 0.5 \tev
$;~\refmark{\chanowitz} and since the
heavy fermions are assumed not to be produced,
it follows that the expressions are relevant
for $ s $ below $1 \tev $. In this range corrections of $ \sim 10 \% $
are possible. These results are of course in agreement with the
estimates of section 4.
Note, however, that it would be very misleading to say that the
corrections are $ O ( s / \mw^2  ) $ (rather than $ \alpha s / 16 \pi^2
m_W^2 $) since this is $ \sim 150 $;
the factors of $g$ and $ 4 \pi $ are very important for any reliable
quantitative estimate.

More insight can be gained by considering the scattering of longitudinally
polarized $W$ vector bosons. At energies such that $ s , \mh^2
\gg \mw^2 , \mz^2 $ the amplitude is~\refmark{\weinberg,
\changeogol, \herrero} (for a review see Ref.~\bargerandphillips)
$$ \acal ( W^+_L W^-_L \rightarrow W^+_L W^-_L )
 = - \sqrt{2} \; G_F \;\mh^2 \left[ { s \over s - \mh^2 }
+ { t \over t - \mh^2 } \right] {\buildrel \mh^2 \gg s \over
\longrightarrow } -           \sqrt{2} G_F u .
\eqn\eq $$ The corresponding $ \ell = 0  $ partial wave is given by
$$ a_0 \simeq { G_F \over 16 \sqrt{2} \; \pi } s, \qquad
\mh^2 \gg s \gg \mz^2 \eqn\eq $$
which grows with $s$ and will
appear to violate unitarity provided $ s \sim
16 \sqrt{2} \; \pi / G_F$ lies in the allowed range for $s$; at  energies
$ \gesim \mh $, however, $a_0$  will go to a constant.

To examine whether tree level unitarity is in fact violated note that,
for the above computations to be reliable, perturbation theory
must be valid. This requires that the Higgs self coupling
be smaller than $ \sim 16 \pi^2 $; so that
$ \mh \lesim 4 \pi v $, where $v$ is the scalar \vev. The
region where $ a_0$ grows linearly with $s$ is bounded by $ s \lesim
\mh^2 $ which, using the above limit for $ \mh $ implies
$ a_0  \ll \pi/2 $ so that unitarity violations are in fact
not generated.
If the Higgs mass is pushed beyond the above perturbative
limit, unitarity is apparently violated, but this
is just a signal that the higher order loop corrections are
as important as the tree level contributions to $ a_0 $: the
scalar sector becomes strongly coupled.~\refmark{\chiral}

\chapter{Radiative Corrections.}

In this section I will review the divergence structure of effective
theories and show that these are renormalizable theories.
The relationship of renormalizability to naturality, gauge invariance
and unitarity will be investigated.

\section{{\bfit{Renormalizability.}}}

Effective theories, just like ``ordinary'' theories can be used
in perturbative calculations; and just as in ordinary field theories
divergences are encountered in loop calculations.
Thus the model requires, as a first step,
regularization; we will use the language of dimensional regularization
since it is guaranteed to preserve the gauge invariance of such importance
in practical calculations. This approach has the defect of hiding
power divergences, but these can be recovered by
considering the model in fewer dimensions.~\refmark{\veltmaniruv}

When studying the perturbative expansion of an observable in an
effective theory a tower of divergences is  obtained whose
structure is more
complicated than the one found in usual renormalizable theories.
These divergences have, however, two important properties: {\it(i)} they
correspond to local interactions; and {\it(ii)} these local interactions
respect all the symmetries of the theory,
assuming that they are preserved
in the (dimensionally or otherwise) regularized theory.

The first property can be proved just as in the usual renormalizable
theories.~\refmark{\kennedy,\collins} Consider any graph $G$ depending on
some external momenta,
collectively denoted by $p$: $ G = G ( p ) $; if we now take sufficient
$p$ derivatives of $G$, the result will be a finite integral. It follows that
$G$ can be written as a polynomial in $p$ with divergent coefficients
plus a finite part. If we now sum all graphs relevant for a given process,
the result will have the same characteristics. The divergent
terms, being polynomial in the external momenta, can be associated with
a set of local operators satisfying the symmetries of the model.
These terms
can be absorbed in renormalizing the coefficients of
$ \leff $ since, by definition,
the effective lagrangian already contains all such operators.
Thus effective theories {\it are} renormalizable. This fact has been used
repeatedly in the chiral approach to strong interactions;~\refmark{\weinberg,
\gl, \pich}
in the \ndec,~\refmark{\longhitano, \escribano, \derujula}
and in the \ndec.~\refmark{\derujula, \arzti, \escribano}

This argument has another application. The presence of effective operator
insertions in loops implies that the effective lagrangian parametrization
is being used at momentum transfers much larger than $ \Lambda $,
and this may appear inconsistent.
To justify the  results obtained in this manner~\refmark{\arzti}
note that, as mentioned
above, taking a sufficient number of derivatives any graph $G$ can be
rendered finite; the resulting integral has then a cut-off equal
to the largest mass in the loop~\refmark{\collins} and this scale is
$ \ll \Lambda $. Therefore the effective
operators can be inserted safely in $ \partial_p^n G( p ) $ for
sufficiently large $n$. Integrating then yields the above
mentioned polynomial ambiguity which is dealt with in the same manner.

The divergences in an effective theory have the same properties
as those arising in an ordinary renormalizable
lagrangian.~\refmark{\londonburgessi}
For gauge theories there are power divergences associated
only with unprotected masses and super-renormalizable
couplings, while the logarithmic divergences
determine the renormalization group running of the couplings in the
theory. Note that working to a fixed order in the loop expansion
and ordering the terms in $ \leff $ according to their index (section
3), insures that to this number of loops the renormalization
group equations involve only a finite number of couplings.

A consequence of the above discussion is that power divergences cannot
be ascribed any phenomenological significance; they are relevant only
for the naturality of certain scalar masses. Note also
that, as a consequence of gauge invariance, the  vector boson masses
can be kept naturally below $ \Lambda $ (see section 7.3).

\section{{\bfit{A simple example.}}}

As a simple example of a loop computation I will consider the operator
$$ \ocal_{ \phi B } = \half \left( \phi^\dagger \phi
- \half v^2 \right) \left(
B_{ \mu \nu } B^{ \mu \nu } \right) \eqn\eq $$
in the \dec, and study its effects on
the $Z$ vector boson vacuum polarization. This calculation is done
to illustrate how loop corrections and renormalization are carried out
within the effective-lagrangian formalism. For more physically relevant
loop calculations see Refs.~\arzti, \hernandez, \roy.

Since this calculation involves a single insertion of a dimension six
operator the results of section 3 insure that all divergences
can be absorbed in renormalizing the terms of dimension $ \le 6 $ in
$ \leff $.

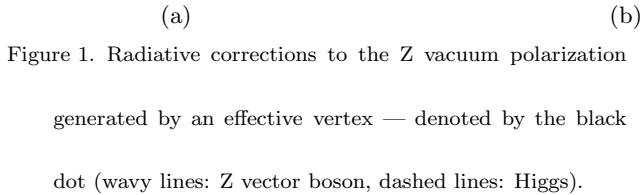
\setbox3=\vbox to 3 truein{\epsfysize=4.5 truein\epsfbox[0 0 612 792]{f1f.ps}}
\centerline{\box3}
\vskip-20pt
\line{{\ninerm \kern 142pt (a) \kern 155pt (b)\hfill}}

\medskip

\centerline{\vbox to .5 in{\hsize=3 in {\eightrm
\iitem{Figure 1.} Radiative
corrections to the Z vacuum polarization generated by an
effective vertex --- denoted by the black dot (wavy lines: Z vector boson,
dashed lines: Higgs).}}}

\bigskip
\medskip

Expanding the dimension six operator yields $$ \ocal_{ \phi B } =
\half \sw^2 Z_{ \mu \nu }^2 \left( \half H^2 + v H
\right) + \cdots ,
\eqn\eq $$ where $ v \simeq 246 \gev $ and
the dots indicate terms that do not contribute to
the calculation at hand (to be done in the unitary gauge).
The lagrangian I will consider is
$$ \lcal \lowti{St. Model}  +
{ \alpha_{ \phi B } \; g'{}^2
\over \Lambda^2 } \ocal_{ \phi B } . \eqn\leffophib $$
The relevant graphs are presented in
\FIG\loopcalc{Loop corrections to the $Z$ vacuum polarization
generated by the operator $ \ocal_{ \phi B } $} figure \loopcalc.
The first diagrams give, in dimensional regularization,
$$ \hbox{Fig.\ \loopcalc  a} = { i \alpha_{ \phi B } \; ( g' \sw )^2\;
m_H^2 \over 8 \pi^2 \Lambda^2 } \left( C_{ U V } +1 - \ln m_H^2
\right) \left( p^2 g^{ \alpha \beta } - p^\alpha p^\beta \right) ,
\eqn\eq $$
where $ C_{ U V } = 2/ ( 4 - n ) -\gamma_E + \ln 4 \pi $ ($
\gamma_E = $Euler's constant and $n$ is the dimension of space time).
The second  graphs give
$$ \hbox{Fig.\ \loopcalc  b} = - { i \alpha_{ \phi B } \; ( g' \sw )^2 \;
m_Z^2 \over 4 \pi^2 \Lambda^2 } \left( C_{ U V } - \ical
\right) \left( p^2 g^{ \alpha \beta } - p^\alpha p^\beta \right) .
\eqn\eq $$ The integral $ \ical $ is defined by $$ \ical = 2 \int_0^1 dx \;
x \; \ln \left[ x m_H^2 + ( 1 - x ) m_Z^2 -x (1 - x ) p^2 \right] .
\eqn\divophib $$
Therefore the contribution to the $Z$ vacuum polarization, which I
denote by $ \delta \Pi_Z ^{ \alpha \beta } $ is transverse, with
$ \delta \Pi_Z ^{ \alpha \beta } = \delta \Pi_Z ( g^{ \alpha \beta }
- p^\alpha p^\beta /p^2 ) $, where $$ \delta \Pi_Z  =
{ \alpha_{ \phi B }  \; ( g' \sw )^2 \;
( m_H^2 - 2 m_Z^2 ) p^2 \over 8 \pi^2 \Lambda^2 }
C_{ U V } + \hbox{finite} . \eqn\eq $$

This divergence can be absorbed in a redefinition of $ \alpha_{ \phi B } $
itself: replacing (the dots indicate higher loop corrections)
$$ \alpha_{ \phi B } \rightarrow
\alpha_{ \phi B } \up 0 = \alpha_{ \phi B } \left[
1 + { 2 m_Z^2 - m_H^2\over 8 \pi^2 v^2 } C_{ U V } + \cdots \right]
\eqn\eq $$ in \leffophib\ cancels the divergence in \divophib.
With this ($\overline{MS}$) renormalization prescription,
and taking for simplicity the
the limit where $ m_H^2 \gg m_Z^2 , p^2 $, I obtain
$$ \delta \Pi_Z\up{\hbox{ren}}  \simeq { \alpha_{ \phi B } ( g' \sw )^2
m_H^2 p^2 \over 8 \pi^2 \Lambda^2 } \left[
1 - \ln { m_H^2 \over \kappa^2 }
+ { 2 m_Z^2 \over m_H^2 } \left( \ln { m_H^2 \over
\kappa^2 } - \half \right) \right] , \eqn\finophib $$ where $ \kappa $ is the
renormalization scale.

It is now natural to ask whether all divergences generated
by $ \ocal_{ \phi B } $ can be absorbed in redefining its
coefficient. This is in fact not the case; for example the
four Higgs Green's function gets a divergent contribution
corresponding to the interaction $ H^2 \square H^2 $. The
full set of divergences will not be described here, I
merely point out that they can all be absorbed in the
coefficients of $ \ocal_{ \phi B } $ or of the operators
$ ( \phi^\dagger \phi - v^2/2 ) | D_\mu \phi |^2 $,
$ ( \phi^\dagger \phi) \square (\phi^\dagger \phi) $
and $ B^{ \mu \nu } ( D_\mu \phi)^\dagger (D_\nu \phi ) $.

This example illustrates the claims made above: the radiative corrections
with an effective lagrangian can be carried out in the same manner as
for ``usual'' lagrangians. The divergences encountered  correspond to
local operators
and therefore can
be absorbed in a renormalization of the existing effective lagrangian
coefficients. In this case
the operators which are renormalized
have the same (or lower) index as $ \ocal_{ \phi B }
$ since the
calculation is done in the \dec, see section 3.
{}From the expression for the counter-terms, the relevant
beta functions can be derived and the running effective couplings
can be obtained. Thus the whole renormalization program~\refmark{\collins}
can be
carried over into the effective-lagrangian
formalism without conceptual difficulty.

Similar calculations can be done in the \ndec, see, for example,
Refs.~\appelquist, \bernard, \longhitano, \arzti. In particular
the first three references explicitly demonstrate that the
divergent terms satisfy the symmetries of the model and hence can
be absorbed in a redefinition of the effective lagrangian coefficients.

When considering quantitative predictions it should not be forgotten
that many processes receive tree-level contributions from $ \leff $.
Tree level contributions
must occur whenever the loop contributions diverge, otherwise these
divergences would not be absorbable in the redefinition of the
coefficients of the effective lagrangian. The converse is not true:
some loops may be accidentally finite, such as the $ \ocal_W $
(see \ow)
contributions to the anomalous magnetic moment of the
muon,~\refmark{\marciano, \arzti}
and still there can be a tree-level operator.
This point is quantitatively important since loop contributions
such as \finophib\ are suppressed by coupling constants and factors of
$ 1 /16 \pi^2 $. See section 8.2 for an example.

In some cases, however,
the tree-level operator contributing to a certain process
is forbidden by a symmetry. It then follows that the
loop contributions must be finite, even if they involve
higher dimension operators.  An example of this situation
is the case where one looks at effective lagrangian corrections for
the two-Higgs-doublet version of the \sm~\refmark{\gunionetal}
and considers the
process $ a_0 \rightarrow \gamma \gamma $, where $ a_0 $ is the CP-odd
scalar of this model.~\refmark{\perez} If the usual
discrete symmetry is imposed to
avoid flavour changing neutral currents,~\refmark{\gunionetal}
then it is easy to see that
there are no tree level contributions from any operator of dimension
six (though they do occur in higher dimensional operators). It follows
that the calculation of this reaction including operators of
dimension six and lower will be finite, which is  verified
explicitly.

\section{{\bfit{Radiative corrections and gauge invariance.}}}

Gauge invariance has been shown to be a very useful principle, but
it is conceivable that this happens to be a low-energy effective
symmetry which is not respected by the heavy physics.~\footnote{l}{Note
that by ``heavy'' I tacitly understand that the scale $ \Lambda $
is at most a few tens of \tev, the case where $ \Lambda $ is the Plank
scale~\refmark{\nielsen} is not studied here.} If true this would imply that
there would be deviations from the gauge invariant couplings generated
by the heavy physics; these coprolites could then produce measurable
effects at various experiments.
I will argue that this is very unlikely based on a an argument originally
put forth by Veltman.~\refmark{\veltmaniruv} The argument is based
on the observation that only when
gauge bosons have Yang-Mills type interactions will the
so-called ``delicate gauge cancellations'' are present. Any deviation
from this type of interaction will ruin this balance with disastrous
consequences.

Consider a model where the vector
bosons have a mass of order $M$ and consider the radiative
corrections to the triple vector boson couplings and to the
fermion-anti fermion-gauge boson couplings. Schematically a triple
vector boson coupling has the form  $ g W^3 p $,
where $W$ denotes the vector boson field, $g$
the corresponding gauge coupling constant, and $p$ a generic
momentum, produced by a derivative which must be present in such
a vertex. The
fermion coupling will be of the form $ g_f \bar \psi \not \! \! W \psi $,
where $ g_f$ denotes the corresponding coupling constant.
Since this is not a gauge theory ($W$ couplings are not
precisely those of the Yang-Mills type but have small corrections
generated by the heavy physics), the only consistent vector boson
propagator has the form $ ( g_{ \mu \nu } - p_\mu p_\nu /M^2 )
/ ( p^2 - M^2 +i \epsilon ) $.

Graph 2.a below
gives a correction of order $ \Lambda^6 g^2/ ( 16 \pi^2 M^4  ) $
to $M^2$ which should, by naturalness,
be $ \lesim M^2 $; the largest allowed value of $ \Lambda $
corresponds to $ M^2 \sim \Lambda^6 g^2/ ( 16 \pi^2 M^4  ) $. This
implies $$ { g \over 4 \pi }
\sim { M^3 \over \Lambda^3 } .
\eqn\eq $$ Similarly, graph b in Fig. 2
gives a correction to the coupling $ g_f $
of order $ g_f^3 \Lambda^2 / ( 16 \pi^2 M^2 ) $, which can at most be $ \sim
g_f $; this requires
$$ { g_f \over 4 \pi } \sim { M \over \Lambda } .
\eqn\eq $$ For a theory which also satisfies $ g \sim g_f $
(as usually imposed) the above naturality arguments imply $ M \sim \Lambda $
and $ g \sim g_f \sim 4 \pi $. Therefore the theory
is strongly coupled. More important, however, is the fact that radiative
corrections will shift $M$ to the cutoff:
the vector bosons do not belong to the low energy theory at all.
Although this is far from being a rigorous argument it does show that
a very careful cancellation of divergences is required if the masses of
the vector bosons are to be kept naturally light.

\setbox1=\vbox to 2 in{\epsfysize=11in\epsfbox[100 -100 712 692]{f2af.ps}}
\setbox2=\vbox to 2 in{\epsfysize=11in\epsfbox[150 -110 762 682]{f2bf.ps}}

\line{\hfill \box1 \kern-5in \box2 \hfill}
\vskip -.5in
\line{{\eightrm  \kern 1.45 in (a)  \kern 2.3 in  (b)\hfill}}

\bigskip

\centerline{\vbox to .2 in{\hsize=3 in {\eightrm
\iitem{Figure 2.} Radiative
corrections for a non-gauge theory (curly lines: vector bosons,
solid lines: fermions).}}}

\bigskip
\medskip

Thus the assumption that
whatever generates the \sm\ as a low
energy effective lagrangian can
violate gauge invariance above a certain scale and also
leave some gauge variant remnants whose effects can be of relevance to
any present or near-future experiments, is inconsistent with
the requirements that the $W$ mass is significantly below this scale, and
with the measurements $ g_f , g \sim 0.65 $.

These arguments in favor of explicit gauge invariance
apparently contradict the previous statements to the effect that any
theory can be rendered gauge invariant (sect. 2).
To understand how these points are reconciled,
recall first that possible contradiction arises only
in the non-linear realization
of the symmetry (else there are no auxiliary
unitary fields present and the
classical lagrangian must be manifestly gauge invariant).

As mentioned in section 4.1 the
corrections to $M$ (obtained by setting $ C = 2 ,  \ B = 0 , \
A' = 2 $ in \chirvertex), are of order
$ g \Lambda / ( 4 \pi ) $. Moreover $ g_f = g $, which follows
from the fact that gauge fields appear always inside
a covariant derivative. These results imply that
naturality is perfectly consistent for a theory which has
been rendered gauge invariant using the Stuckelberg trick, provided the
gauge boson mass is related to the cutoff
in the above manner; identifying $ \Lambda = 4 \pi v $
yields the familiar result $ M \sim g v $. The difference between this
calculation and the previous one lies in the
intrinsic properties of gauge-invariant
theories: only in these models the vector-boson
propagators have a reasonable high-energy behaviour,
and this leads to much milder constraints. In contrast, when gauge
invariance is not present, the only consistent
vector boson propagators
do not drop off at large momenta, and this leads to unacceptably
large corrections to the vector boson masses; the theory,
moreover, must also
be strongly coupled. Both these results are unacceptable for the
electroweak sector.

An important conclusion to be drawn from these
arguments it that the fact that we observe
the $W$ and $Z$ bosons with masses significantly
below the Fermi scale gives
a very strong argument for imposing gauge invariance in the lagrangian
(be it via the Stuckelberg trick or directly).

\section{{\bfit{Anomalies.}}}

The presence of new fermion-gauge boson couplings in $ \leff $
raises the possibility that new anomalies are generated.
This is most easily discussed by using a high derivative
regularization.~\refmark{\anom} The basic
idea is to replace
$$ \bar \psi i \not \! \! D \psi \rightarrow
\bar \psi  \left\{ i \not \! \! D \left[ 1 + \left( - D^2 / M^2
\right)^n \right] \right\} \psi ,
\eqn\eq $$ so that the propagator drops off
as $ 1/ p^{ 2 n + 1 } $ at large momentum $p$. Using this regularization
method simple power counting shows that
all graphs but the usual triangle
(with the minimal substitution
fermion-gauge boson vertices) ones are well defined
whenever $n$ is sufficiently large. This implies that the axial currents
would be well defined in the regulated theory,
were it not for these graphs, which
correspond to the usual anomalous diagrams. It follows
that there are no new anomalies
generated by the higher dimensional operators.

\chapter{Applications.}

There has been recently a surge of papers applying the effective
lagrangian parametrization to various processes covering a wide
variety of subjects, from
the recent observation of the $ B \rightarrow K^* \gamma $ decay to
various reactions in $ \gamma \gamma $ (back scattered laser)
colliders. It is impossible to review all of these papers in any
detail and so I am forced to make a selection of various cases which
are illustrative; I will, however, mention related results when
appropriate.

The results will be organized as follows.
Given an experiment I will present the operators whose effects have
been studied together with the bounds on their coefficients
(either real bounds form current data or expected sensitivity for
future experiments).
In the discussion
below the reader will often find the phrase ``the sensitivity limits are''
followed by an equation of the type $|${\it quantity}$| < ${\it bound}.
This should be understood to mean that the experiment in question
will be insensitive to ``{\it quantity}'' if its values happen to lie
in the interval $ ( - {\it bound} , + { \it bound} )$.

Sections 8.1 --- 8.11 deal with specific experiments. the results
are summarized in the tables of section 8.12. The reader not interested
in a specific case may proceed to that section directly.

For the most of this section the coefficients
will be assumed to take their natural sizes (see section 4) which
entails, aside from possible positive or negative powers of $ 4 \pi $,
the presence of gauge coupling constants whenever a gauge field is
present. In this respect there are two different
situations which lead, in general, to different natural values for the
coefficients. In the \ndec\ the relevant estimates are obtained from
\chirvertex. For the \dec, following the discussion of section 4.2,
I will assume that the underlying physics is weakly coupled, hence
the coefficients are obtained by determining
whether an operator is tree or loop generated (section 4.2);
the coupling constants will be assumed to be $ \sim 1 $.

When considering the contributions from $ \leff $ to an observable it
is often found that many terms contribute. The modification to
the \sm\ prediction will be then proportional to a linear combination of
the coefficients of the contributing operators. I will assume that
there are no (significant) cancellations between these coefficients
in order to derive bounds on them. Alternatively one can say that this
linear combination defines a unique operator which contributes to
the observable of interest, and that the coefficient of this operator
can be estimated following the arguments of section 4.

In the \dec\ the operators for which I will present limits are
$$ \eqalign{
\lcal_{ l W } &= \ila g \; \alpha_{ l W }
\left( \bar \ell_l \; \sigma^{ \mu \nu } \; \sigma_I
l_R \right) \phi \; W_{ \mu \nu}^I ; \cr
\lcal_{ l B } &= \ila g' \; \alpha_{ l B }
\left( \bar \ell_l \; \sigma^{ \mu \nu }
l_R \right) \phi \; B_{ \mu \nu} ; \cr
\lcal_W &= \ila g^3 \; \alpha_W \epsilon_{ I J K } W_{ \mu \nu }^I
W_{ \nu \rho }^J W_{ \rho \mu }^K ; \cr
\lcal_{ \tilde W }  &= \ila g^3 \; \alpha_{ \tilde W }
\epsilon_{ I J K } \tilde W_{ \mu \nu }^I
W_{ \nu \rho }^J W_{ \rho \mu }^K ; \cr
\lcal_{ W B } &= \ila g g' \; \alpha_{ W B }
\phi^\dagger \sigma_I \phi W_{ \mu \nu }^I B_{ \mu \nu } ; \cr
\lcal_{ \tilde W B } &= \ila g g' \; \alpha_{ \tilde W B }
\phi^\dagger \sigma_I \phi \tilde W_{ \mu \nu }^I B_{ \mu \nu } ; \cr
\lcal_{ 4 \psi ; vec} &= \ila \alpha_{ 4 \psi ; vec}
\sum_{f =  q , \ell_l } \left( \bar q \gamma^\mu q \right)
\left( \bar \ell_l \gamma_\mu \ell_l \right) ; \cr
\lcal_{ 4 \psi ; sc} &= \ila \alpha_{ 4 \psi ; sc}
\left[ \left( \bar \ell e \right) \epsilon \left( \bar q u \right)
+ 2 \left( \bar \ell u \right) \epsilon \left( \bar q e \right)
\right] ; \cr
\lcal_{ \phi f } \up 3 &= \ila \alpha \up 3_{ \phi f }
\left[ \sum_l \left( \bar \ell \gamma^\mu \sigma_I \ell \right)
\left( \phi^\dagger \sigma_I D_\mu \phi \right) +
\sum_q \left( \bar q \gamma^\mu \sigma_I q \right)
\left( \phi^\dagger \sigma_I D_\mu \phi \right) \right] ; \cr
\lcal_{ \phi W } &= \ila { g^2 \; \alpha_{ \phi W} \over 2 }
( \phi^\dagger \phi) \left( W_{ \mu \nu } ^I \right)^2 ; \cr
\lcal_{ \phi \tilde G } &= \ila { g_s^2 \; \alpha_{ \phi \tilde G }
\over 2 } \left( \phi^\dagger \phi \right) \tilde G_{ \mu \nu }^A G^A{} ^{
\mu \nu } ; \cr
\lcal_\phi \up 1 &= \ila \alpha_\phi \up 1 \left( \phi^\dagger \phi \right)
\left( D_\mu \phi^\dagger D^\mu \phi \right) ; \qquad
\lcal_\phi \up 3 = \ila \alpha_\phi \up 3 \left| \phi^\dagger D_\mu \phi
\right|^2 ; \cr
} \eqn\decops $$ \con. I have assumed that the couplings in $ \lcal_{ 4 \psi
 ; vec } $ are all identical; this is done for simplicity only, for a
more detailed analysis see, for example Refs.~\ehlq\ and \ruckl.

The natural order of magnitude for the above coefficients is
$$ \eqalign{
 \alpha_{ 4 \psi ; vec, sc } ,
\alpha_{ \phi f } \up 3 , \alpha_\phi \up{ 1,  3 } & \sim 1 ; \cr
 \alpha_{ l W } , \alpha_{ l B } ,
\alpha_{ W B } , \alpha_{ \tilde W B } , \alpha_{ \phi W } , \alpha_{ \phi
\tilde G } ,\alpha_W , \alpha_{ \tilde W } &
\sim \inv{ 16 \pi^2 } . \cr } \eqn\wicest $$

In the \ndec\ the operators considered are
$$ \eqalign{
\lcal_1' &= \quarter \beta_1 g^2 v^2 \left\{ \tr \left( \sigma_3
U^\dagger D_\mu U \right) \right\} ; \cr
\lcal_1 &= \quarter \alpha_1 \; g g' \; B^{ \mu \nu }
W_{ \mu \nu }^I \;  \left( U \sigma_3 U^\dagger \sigma_I \right) ; \cr
\lcal_2 &= i \alpha_2 \; g' \; B^{ \mu \nu } \tr \left( \sigma_3
D_\nu U^\dagger \; D_\mu U \right) ; \cr
\lcal_3 &= i \alpha_3 \; g \;  W^I_{ \mu \nu } \tr \left( \sigma_I
D^\nu U^\dagger \; D^\mu U \right) ; \cr
\lcal_4 &= \alpha_4 \left\{ \tr \left(  D_\nu U^\dagger \; D_\mu U \right)
\right\}^2 ; \cr
\lcal_5 &= \alpha_5 \left\{ \tr \left(  D^\mu U^\dagger \; D_\mu U \right)
\right\}^2 ; \cr
\lcal_8 &= \inv{16}  \alpha_8 g^2 \; \left\{
W_{ \mu \nu }^I \tr \left( U^\dagger \sigma_3 U
\sigma_I \right) \right\}^2 \cr
\lcal_{ 11 } &= \alpha_{ 11 } \; g \; \tilde W_{ \mu \nu } ^I
\left[ \tr \left( \sigma_3 U^\dagger D^\mu U \right) \right]
\left[ \tr \left( U^\dagger \sigma_I D^\nu U \right) \right] ; \cr
} \eqn\ndecops $$ where $D_\mu U = \partial_\mu U
- { i \over 2 } g \sigma_I
W_\mu^I U - { i \over 2 } g' B_\mu U \sigma_3  $. The operators
$ \lcal_{ 4 \psi ; vec, sc } $ also appear in the \ndec\ to the order
considered here (index$\le 2 $). The estimates from section 4 imply
$ \alpha_i \sim 1/ 16 \pi^2 $. We also would have $ \beta_1 \sim 1 $
but, as mentioned before, there are extra suppression factors
(of unknown origin) which require this constant to be $ \lesim 1 \% $
(which coincidentally is of the same order as the $ \alpha_i $).
The terms $ \lcal_{ 4 \psi; vec ,sc } $ are also present in the \ndec\
to the order we are working; as discussed in section 4 their coefficients will
be $\sim 16 \pi^2 $, that is, these terms are $ \propto 1 / v^2 $.
In the following I will assume $$ \alpha_i , \beta_1 \sim \inv{ 16 \pi^2 } ;
\qquad \alpha_{ 4 \psi ; vec , sc } \sim 16 \pi^2 .
\eqn\ndecest $$

Two tables will be  given, one where the
sensitivity of the various present and future experiments to $ \Lambda $
when the coefficients of $ \leff $ take their natural values,
and another where the bounds on the coefficients
are derived when $ \Lambda = 1 \tev $. This choice is made for convenience.
It must be remembered that in the \dec\
when the underlying theory is weakly coupled,
there may be contributions arising from tree-level-generated dimension
eight operators which can dominate over the ones presented here
(section 4.2). The
crossover occurs at scales $ \sim 3 \tev $. This can produce
enhancements of the order of $ ( 4 \pi v / \Lambda )^2 \sim 10 $
(when $ \Lambda = 1 \tev $) in the expected
magnitude of the measured effects. On the other hand there may be
small coupling constants that can decrease the contribution by
an order of magnitude or more. The results should be
interpreted keeping these caveats in mind.

Before proceeding to the predictions a comment on the triple vector boson
couplings is needed. The most general Lorentz invariant
lagrangian describing such terms (see, for example, Ref.~\hagiwara)
$$ \eqalign{
\lcal_{ WWV } / g_{ WWV } = &
i g_1^V \left( W_{ \mu \nu } ^\dagger W^\mu V^\nu - \hbox{ h.c. } \right)
- g_4^V
W_\mu^\dagger W_\nu \left( \partial^\mu V^\nu + \partial^\nu V^\mu
\right) \cr &
+ i \kappa_V W_\mu^\dagger W_\nu V^{ \mu \nu }
+ i { \lambda_V \over \mw^2 } W^\dagger_{ \lambda \mu }
W^\mu{}_\nu V^{ \nu \lambda } \cr
&  + i \tilde \kappa_V W^\dagger_\mu W_\nu \tilde V^{ \mu \nu }
+ i { \tilde \lambda_V \over \mw^2 } W^\dagger_{ \lambda \mu } W^\mu
{}_\nu \tilde V^{ \mu \nu } \cr
& + g_5^V \epsilon^{ \mu \nu \rho \sigma } \left(
 W_\mu {\buildrel \leftrightarrow \over \partial_\rho } W_\nu \right)
V_\sigma \cr } \eqn\tbv $$ (where terms
proportional to $ \partial \cdot W $ and $ \partial \cdot V $ are
ignored). $V$ denotes either the photon or the $Z$ field; in
the first case only terms in compliance with electromagnetic gauge
invariance are retained. It is assumed (without loss of generality)
that $ g_{ WW \gamma } = -e , \ g_{ WWZ }  = - e \cot \thew $.

In the literature bounds on the couplings $ \kappa_V , \lambda_V $,
etc. are often found. It is important to remember, however, that
the use of a consistent gauge-invariant
effective lagrangian expansion imposes
severe constraints among these couplings. In the \dec
$$ \eqalign{
 \lambda_\gamma = \lambda_Z = { 6 m_W^2 \; g^2 \over \Lambda^2 } \alpha_W
; \quad & \quad
\tilde \lambda_\gamma =
\tilde \lambda_Z = { 6 m_W^2 \; g^2 \over \Lambda^2 } \alpha_{ \tilde W } \cr
 \kappa_\gamma - 1 = \kappa_Z - 1 = { 4 m_W^2 \over
\Lambda^2 } \alpha_{ W B } ; \quad & \quad
\tilde \kappa_\gamma = \tilde \kappa_Z = { 4 m_W^2 \over
\Lambda^2 } \alpha_{ \tilde W B } . \cr } \eqn\dectransl $$

In the \ndec, and in the notation of appendix B,
the corresponding relations for the CP conserving operators
are~\refmark{\appelquistwu}
$$ \eqalign{
g_1^Z - 1 &= { g^2 \over \ctw } \beta_1 + { g^2 \tw^2 \over \ctw } \alpha_1
+ { g^2 \over \cw^2 } \alpha_3 ; \cr
\kappa_Z -1 &= { g^2 \over \ctw } \beta_1 + { g^2 \tw^2 \over \ctw } \alpha_1
+ { g^2 \tw^2 } ( \alpha_1 - \alpha_2 ) + g^2 ( \alpha_3 -
\alpha_8 + \alpha_9 ) ; \cr
\kappa_\gamma - 1 &= g^2 ( - \alpha_1 + \alpha_2 + \alpha_3 - \alpha_8
+ \alpha_9 ) ; \cr
g_5^Z &= g^2 \tw^2 \; \alpha_{ 11 } ; \cr
g_5^\gamma &= g_1^\gamma - 1 = \lambda_Z = \lambda_\gamma = 0 .
\cr } \eqn\ndectransl $$
Some of these constants can be $ \sim 0.01 $ if all contributions
add constructively. If this is not the case they are $ \sim 0.003 $.

In the following discussion I will continuously refer to these relations.
It must be pointed out, however, that \dectransl\ and \ndectransl\ apply
only for the operator basis chosen. If, for example, the operator
$ B^{ \mu \nu } D_\mu \phi^\dagger D_\nu \phi $ is added,
its coefficient shifts $ \kappa_Z $ away from $ \kappa_\gamma $.
This operator is redundant (in the sense of section 5)
and hence its addition cannot affect any observable: its
effect can be absorbed in a redefinition of the coefficients of
the \sm\ parameters and other operators. I will
consistently use the basis presented in Ref.~\bw\ for the following
analysis.

I will not present all the references to the experimental papers
but instead refer the reader to the papers cited for their
precise sources.

\section{{\bfit{Low energy results.}}}

In this sub-section I present some of the more spectacular
bounds on $ \Lambda $ derived from low energy effects.
This is very far from a complete list, the reader is referred to
the extensive study in Ref.~\bw\ of the low energy effects
of the dimension six operators in the \dec.

\item{\bullet} {\it Neutron dipole moment.}
The term $ \lcal_ {\phi \tilde G } $
generates a contribution to the CP violating $ \theta $ parameter,
which is $ < 10^{ - 9 } $;~\refmark{\partdatbook} this implies $$ { | \alpha_{
\phi
\tilde G } | \over \lat^2 } < 10^{ - 10 } . \eqn\eq $$ Using \wicest\
this implies $ \lat > 8000 $. This presents a bound on the scale at which
processes contributing to the neutron dipole moment occur.

\item{\bullet} {\it $ K_L -  K_S $ mass difference.} The contributions
to this quantity depend on operators which violate flavor conservation;
the bounds on $ \Lambda $ should then be identified as constraints
on the scale at which flavor-changing neutral currents are generated.
In this example only I will
include more than one generation of fermions. The terms in $ \leff $
which I will consider is $$ \lcal_{FCNC} = { \alpha_{ FCNC } \over 2
\Lambda^2 } \left( \bar q_1 \gamma_\mu q_2 \right)^2 \eqn\eq $$
where $ q_1 $ denotes the up-down quark doublet and $ q_2 $ the
charm-strange quark doublet. The contribution to the neutral kaon
mass difference requires $$ { | \alpha_{ FCNC } | \over \lat^2 }
< 7 \times 10^{ - 7 } . \eqn\eq $$ This limit corresponds to
$ \lat > 1200 $ when $ | \alpha_{FCNC} | = 1 $.

The above bounds are special in that they are obtained for operators
which are associated with certain very much suppressed processes within the
\sm. Thus the corresponding scales might easily be very different
from the $ \Lambda $ appearing in other operators.~\footnote{m}{This same
comment applies to $ \lcal_{ 4 \psi ; sc } $ in \decops.} For other
operators the bounds obtained in Ref.~\bw\ are comparable or weaker than
the ones discussed below (albeit using some operators not included in
\decops).

\section{{\bfit{AGS821.}}}

The Brookhaven experiment AGS821~\refmark{\ags}is
expected to measure $ a_\mu =
(g_\mu - 2)/2 $ for the
muon to an accuracy of $ 4 \times 10^{ - 1 0 } $. This can be used
to determine the expected sensitivity to the anomalous magnetic
moment coupling effective operators~\refmark{\arzti} $
\ocal_{ \mu W } $ and $ \ocal_{ \mu B } $ in \decops.
The parameter values to which this experiment will not be sensitive lie
in the region $$ { | \alpha_{ \mu W } - \alpha_{ \mu B } |
\over \lat^2 } < 1.1 \times 10^{ -5 } .\eqn\eq $$

In most models, chiral symmetry implies that
the coefficients $ \alpha_{ \mu W , \mu B } $
are suppressed by the muon Yukawa coupling $ y_\mu $. This is
suggested naturally
by the smallness of the muon mass (but can be avoided in
some models, see Ref.~\arzti). Then, assuming no cancellations, the
above limits imply, for the \dec, $ \lat > 0.5 $.
For the \ndec\ the same results hold provided the substitution $
\Lambda \rightarrow v \simeq 0.246 \tev $ is made. The CERN
data~\refmark{\partdatbook}
strongly favors the presence of the $ y_\mu $ suppression factor.
In the tables this suppression factor is assumed.

This measurement has also been used to put a limit on the effective
couplings among three vector bosons (see Ref.~\marciano\ for a review).
With the natural
value for the coefficients the contributions from the corresponding
operators are unobservable unless $ \Lambda $ is a few \gev\ (in the
\dec). For a full discussion see Ref.~\arzti.

\section{{\bfit{CLEO.}}}

The measurement of the branching ratio $ B \rightarrow K^* \gamma
$~\refmark{\cleo} has been used to impose
bounds on the effective couplings among
triple vector boson vertices.
The bounds derived from the CLEO measurement are~\refmark{\hemackellar}
$ -0.13 \lesim 1 - \kappa_\gamma \lesim 0.75 $,
$ -2.2 \lesim \lambda_\gamma \lesim 0.4 $,
$ | \tilde \kappa_\gamma | \lesim 0.32 $,
$ | \tilde \lambda_\gamma | \lesim 0.93 $; when the top mass is
$ m \lowti{ top } = 150 \gev $. Weaker bounds
($ | \lambda_\gamma | \lesim 10 $ when all other constants in \tbv\
vanish) were obtained in Ref.~\peterson.
This implies, using \dectransl,
$$ \eqalign{ - 5.1 < { \alpha_{ W B } \over \lat^2 } < 29.3
\quad & \quad { | \alpha_{ \tilde W B } | \over \lat^2 } < 12.5 \cr
- 135.6 < { \alpha_{ W } \over \lat^2 } < 24.7
\quad & \quad { | \alpha_{ \tilde W  } | \over \lat^2 } < 57.3 \cr } . $$

In Ref.~\dawson\ the effects of $ \lcal_{ 11 } $ in \ndecops\ on the decays
$ B_s \rightarrow \mu^+ \mu^- $ were studied. The sensitivity limit
is $$ | \alpha_{ 11 } | \lesim 1.8 ; \eqn\eq $$ similar results
are obtained from the decay $ K^+ \rightarrow
\pi^+ \nu \bar \nu $.

\section{{\bfit{HERA.}}}

There have been several studies into the bounds on four fermi operators
that can be obtained by HERA. The possible four fermi interactions
can be understood as being generated via a heavy vector
exchange, $ \lcal_{
4 \psi ; vec} $; or a heavy scalar exchange, $ \lcal_{ 4 \psi ; sc } $.
These are labelled vector and scalar exchange
respectively.~\footnote{n}{All other possibilities are equivalent via a
Fierz transformation.} The coefficients for these terms have large natural
value (see \wicest) which implies good sensitivity to $ \Lambda $.

The vector exchange terms were studied in, for example,
Refs.~\ruckl, \doncheski\
using deep inelastic
polarized $ e^\mp $-nucleon scattering. A typical result
is~\refmark{\doncheski}
$$ { | \alpha_{ 4 \psi; vec} | \over \lat^2 } \lesim 0.4 \eqn\eq $$

The scalar exchanges have been considered in Refs.
\burgesschnitzer, \doncheski,
\wudkahera. These are based on the possibility of using polarized electrons
to probe helicity violating interactions. Low energy data~\refmark{\bw}
restrict these interactions to the form\footnote{o}{Using a Fierz
transform this
can be identified with a tensor exchange.}
$ \lcal_{ 4 \psi ; sc } $ in \decops. For
70\% polarization  the sensitivity is
to $$ { | \alpha_{ 4 \psi ; sc } | \over \lat^2 } < 7 . \eqn\eq $$

The lagrangian \tbv\ has also been studied for this accelerator.
In Ref.~\helbig\ the reaction $ e p \rightarrow \nu \gamma X $ is
used to set the bounds $ | \kappa  - 1 | \lesim 1.4 ; \
| \lambda  | \lesim 1.1 $
(a similar study,~\refmark{\kim} but for five year's integrated luminosity,
improves these limits to
$ | \kappa - 1 | \lesim 0.3 $ and $ | \lambda | \lesim
0.8 $). These are too weak to be of interest in the \ndec; for the \dec\
they imply $$ { | \alpha_W | \over \lat^2 } < 68 ; \qquad { | \alpha_{ W B }
| \over \lat^2 } < 55 . \eqn\eq $$

\section{{\bfit{LEP1.}}}

The wealth of experimental data from LEP1 has been used by several authors
to impose bounds on various operators.

\item{\bullet}{\it Four fermi interactions}.
As in the colliders reviewed above, the four fermion interactions
are very sensitive to new physics. For example, the forward-backward
asymmetry in $ b \bar b $ production generates a limit~\refmark{\langacker}
$$ { | \alpha_{ 4 \psi ; vec } | \over \lat^2 } < 1.9 \eqn\eq $$

\item{\bullet}{\it Measurements with LEP data and $m_W$ as inputs}.
Another good  limit on $ \Lambda $ in the \dec\ is obtained in Ref.
\derujula\ for the coefficient of the operator $ \ocal_\phi \up 1 $
in \decops.~\footnote{p}{The operator
actually considered is $ \ocal_\phi \up 1 -
\half \left[ \partial ( \phi^\dagger \phi ) \right]^2 $;
the second term, however, vanishes in the unitary gauge.}
According to \wicest\ the coefficient of such
an operator is comparatively large. The bounds obtained are $$
-0.132 < { \alpha_\phi \up 1  \over \lat^2 } < 0.876 \eqn\eq $$
This implies
$ \Lambda \gesim 1.5 \tev $.
Similar bounds were also obtained from leptonic four fermion
operator effects. These results were based on the measurement of the $Z$
mass and widths as well as the $W$ mass and the neutrino
cross section ratio.

\item{\bullet}{\it Triple vector boson vertices}.
In a related investigation the authors of
Ref.~\hagiwaraii\ fit the values of several operators to the
data and obtain the bounds $ 0.7 < \kappa_\gamma < 1.7 $ and $ | \lambda_
\gamma | < 0.6 $ which translates into $$ -12 < { \alpha_{ W B } \over
\lat^2 } < 27 ; \qquad { | \alpha_W | \over \lat^2 } < 37 . \eqn\eq $$
These authors also consider bounds on the \ndec\ operators \ndecops,
$$
-0.11 < \beta_1 < 0.02 , \quad
-0.05 < \; \alpha_1 < 0.04 ,\quad
| \alpha_8 | < 0.04  ;
\eqn\eq $$ where all the
couplings are naturally of order $ 1 / 16 \pi^2 $, see \ndecest.

\item{\bullet}{\it $\tau $ anomalous moments}.
The authors of Ref.~\escribano\ use the $ Z \rightarrow \tau^+ \tau^- $ decay
rate together with the CDF measurement of $ m_W $~\refmark{\partdatbook}
and the
ratio of the charged to neutral neutrino cross sections
to impose bounds on the contributions to
$ a_\tau $ (the anomalous magnetic moment)
and $ d_\tau $ (the anomalous electric dipole moment)
generated by $ \ocal_{ \tau B } $ in \decops. To $ 2 \sigma $
they find a bound $$ { | \alpha_{ \tau B } | \over \lat^2 } < 1.1 ,
\eqn\eq $$ with the natural size for the
coupling given in \wicest.

\item{\bullet}{\it Custodial symmetry breaking}.
In Ref.~\derujula\ bounds are obtained on the coefficient of the operators
$ \ocal_{ W B } $ which is essentially the oblique $S$ parameter, see
\decst. The experimental results used~\refmark{\partdatbook} were
the $W$ mass, the partial
$Z$ widths, the leptonic axial $Z$ coupling and
the neutrino cross section ratio.
The $ 2 \sigma $ limit on this coefficient is
$$ -0.6 < { \alpha_{ W B } \over \lat^2 } < 0.7 \eqn\eq $$
Using \wicest\ implies that
LEP1 is sensitive to scales up to $ \sim 100 \gev $ (which
is essentially the value of $ \sqrt{ s } $ for this accelerator).

\item{}In Ref.~\kneur\ breaking of the custodial $ \su2 $ symmetry in the form
of an effective $ j_\mu^3 B^\mu $ coupling ($ j_\mu^I $ is the fermionic
current of weak isospin index $I$). In terms of the parametrization of
Ref.~\bw\ this coupling corresponds to $ \ocal_{ \phi f } \up 3 $ in \decops.
The measurements of
$ m_W $, $ m_Z $ and the weak mixing angle imply $$
{ | \alpha_{ \phi f } \up 3 | \over \lat^2 } \lesim 1 \eqn\eq $$
This term, like the four-fermion interactions, has a relatively large
coefficient and therefore can provide substantial bounds on $ \Lambda $.

Other investigations into the custodial symmetry breaking involve
the $ \rho $ parameter to which I now turn.

\item{\bullet}{\it Oblique parameters}.
The oblique parameters $S$, $T$ and $U$~\refmark{\peskin}
within the \ndec\ are obtained in  Refs.~\golden, \greenstein, \holdomi.
The results, in the notation of appendix B,
are~\refmark{\appelquistwu} $$ S = - 16 \pi \alpha_1 ; \quad T = { 8 \pi \over
\sw^2 } \beta_1 ; \quad U = - 16 \pi \alpha_8 \eqn\eq $$ and are
expected to be of order $ 1 / \pi $.
Estimates of these quantities for several
models are also provided in these references. These expressions
together with
the experimental results~\refmark{\ellis} $ -0.8 < S < 0.18 $,
$ -0.46 < T < 0.22 $ and $ -1.03 < U < 0.81 $ provide the bounds
$$ -0.004 < \alpha_1 < 0.016 ; \quad
-0.004 < \beta_1 < 0.002 ; \quad -0.016 < \alpha_8 < 0.02 .
\eqn\lepndeclimits $$
These limits are already of the same order as the estimates in
\ndecest, yet one more indication of the excellence of LEP data.

\item{}
Similar expressions can be obtained
in the \dec, namely $$ S = 32 \pi
\alpha_{ W B } { v^2 \over \Lambda^2 } ;
\quad T = - { 4 \pi \over \sw^2 } \alpha_\phi \up 3
{ v^2 \over \Lambda^2 } ; \quad U = 0 , \eqn\decstu $$
where only dimension six
operators have been kept. An immediate result is that the parameter $U$
provide a clear differentiation between the \ndec\ and the \dec\
provided the operators of dimension eight in the latter case can be
ignored (\ie\ provided $ \Lambda \gesim 3 \tev $).~\refmark{\wudkamerida}
The above limits on $S$ and $T$ translate into
$$ - 0.13 < { \alpha_{ W B } \over \lat^2 } < 0.03 ; \qquad
-0.06 < { \alpha_\phi \up 3 \over \lat^2 } < 0.13 ,\eqn\eq $$
which put significant bounds on $ \Lambda $: $ \gesim 300 \gev $.

It must be remembered that the expressions \decstu\ depend on the basis
of operators chosen, if another basis is chosen they
are modified (for an example see Ref.~\hagiwaraii);
all discrepancies will disappear when the results are expressed
in terms of observables.

The effects of \tbv\ on the oblique parameters were calculated, for example,
in Ref.~\roy. These authors did not impose the constraints \dectransl\
and so their estimates are relevant for the \ndec\ scenario only. A typical
result
is $ - 1.4  < \kappa_Z - 1 < 0.4 $ which is too loose to generate significant
information.

\item{\bullet}{\it Radiative effects}.
The effects of the operator
$ \ocal_W $ are also considered in Ref.~\derujula. They
affect the measured observables only through radiative corrections
(this is true for all
``blind directions'', see sections 5 and 7 for a discussion of this
type of operators). The
induced shift on the photon, $Z$ and $W$ vacuum polarization tensors
gives the limit
$$
-1.4 < \alpha_{ W  } { m_W^2 \over \Lambda^2 }
\ln { \Lambda^2 \over m_W^2 } < 4.4 ; \eqn\eq $$
where the natural size of the coupling is given in \wicest;
the derived bounds on $ \alpha_W $ are very weak for $ \Lambda > m_W $.

In a related calculation the radiative
effects of other blind operators (section 5)
were studied in Ref.~\hernandez. For example,
using the same observables as above, the effective interaction
$ i \alpha'_{ \phi B } \; g' \;
B^{ \mu \nu } D_\mu \phi^\dagger \; D_\nu \phi $ was
bounded with the result $ | \alpha' _{ \phi B } |
/ \lat^2 < 6 $.  As mentioned at the end of section 5, however,
this blind direction can be given sight by choosing a different basis. In
this case the effects on the $ \rho $ parameter (for example) yield
better limits and with much less effort.

\section{{\bfit{Tevatron.}}}

The best sensitivity to $ \Lambda $ for the Fermilab Tevatron
is obtained from the four-fermion term $ \lcal_{ 4 \psi ; vec} $
in \decops. As an example, in Ref.~\wainer\ the sensitivity limit
derived is $$ { | \alpha_{ 4 \psi ; vec } | \over \lat^2 } < 3 \eqn\eq $$

The investigations concerning the sensitivity of the Fermilab Tevatron
to \tbv\ have been published in Refs.~\hagiwarazep,\bauri. The
predicted results depend, of course,
on the luminosity available. For example, looking at $WZ$ production with
only leptonic decays for the vector bosons gives an expected
bound (from CDF)
of $ | \lambda_Z | < 1.7 $ for an integrated luminosity of $ 4.7 /$pb.
This is improved to $ | \lambda_Z | < 0.4 , \ | \kappa_Z - 1 | < 2 $
for $ 100 / $pb. For $W^+W^-$ production, with only leptonic $W$ decays
selected as the final state gives the bound
$ | \lambda | < 1 , \ | \kappa - 1 | < 1.3 $
with $ 4.7 / $pb of integrated luminosity. Finally, using CDF and
D0 data for $W \gamma $ production at 95\% confidence
level, $ | \lambda_\gamma | < 0.31 , \ | \kappa_\gamma - 1 | < 1.15 $.
Summarizing, the sensitivity of the Tevatron to \tbv\ is
determined by $ | \lambda | \lesim 0.4 , \ | \kappa_Z - 1 | \lesim 1.5
$. This is not very restrictive within the \ndec; for the \dec\ they
correspond to $$ { | \alpha_{ W B } | \over \lat^2 } < 59 ; \qquad
{ | \alpha_{ W } | \over \lat^2 } < 25 ;
\qquad \hbox{(for}\sim 100/\hbox{pb)} . \eqn\eq $$
These are, as mentioned above, predictions. The latest
measured limits from D0~\refmark{\ellison} using $ W \gamma $ production are,
for
$ 15 /$pb,
$| \lambda | \lesim 1.2 , \ | \kappa_Z - 1 | \lesim 2.6 $
corresponding to
$$ { | \alpha_{ W B } | \over \lat^2 } < 102 ; \qquad
{ | \alpha_{ W } \over \lat^2 } < 74 ; \qquad
\hbox{(for}\sim 15/\hbox{pb)}  \eqn\eq $$
The natural size for the coefficients is given in \wicest.

\section{{\bfit{LEP2.}}}

There have been many predictions as to the sensitivity of LEP2 to various
operator coefficients. The CM energy is assumed to be $ 190 \gev $ and the
luminosity  $ 500 / $pb.

The bounds on the four-fermion interaction coefficients are, for the
leptonic final states~\refmark{\barbiellini} $$ { | \alpha_{ 4 \psi ; vec} |
\over \lat^2 } < 0.13 .\eqn\eq $$

The bounds on \tbv\ are derived from differential cross section for
the process $ e^+ e^- \rightarrow W^+ W^- $.~\refmark{\bilenky,
\hagiwara, \hagiwarazep}
The sensitivity limit of LEP2 is given by $ | \kappa_V - 1 | ,
|\lambda_V | , | \tilde \kappa_V | , | \tilde \lambda_V |
\lesim 0.5 $. These
are too large for the \ndec; in the \dec\ they imply the constraints
$$ { | \alpha_{  W } | \over \lat^2 } ,
 { | \alpha_{ \tilde W } | \over \lat^2 } < 31; \qquad
 { | \alpha_{  W B } | \over \lat^2 } ,
 { | \alpha_{ \tilde W B } | \over \lat^2 } < 20 . \eqn\lepii $$
These bounds are derived assuming $ \kappa_\gamma = \kappa_Z $
and $ \lambda_\gamma = \lambda_Z $ which follows from the triple vector
boson interactions derived from \dec\ $ \leff $ (see appendix
A); the limits \lepii\
are considerably weakened if this assumption is
relaxed.~\refmark{\bilenky}

The authors of Ref.~\kneur\ have also investigated the $ j_\mu^3 B^\mu $
couplings in LEP2 (see the subsection on LEP1). Assuming that the
top mass is found at the Tevatron, the expected sensitivity
becomes $$
{ | \alpha_{ \phi f } \up 3 | \over \lat^2 } \lesim 0.06 \eqn\eq $$
which improves the sensitivity to $ \Lambda $ by a factor of four,
to $ 2.5 \tev $ in the weak coupling case, and to $ 10 \tev $ in the
strong coupling case.

The \ndec\ has also been studied for LEP2. Reference \holdomii\ considers
the effects of the terms $ \lcal_{ 1 , 2 , 3} $
of \ndecops\ (assuming $ \alpha_2 = \alpha_3 $)
on $W^+W^-$  production. No significant deviations from the
\sm\ were found even if the $ \alpha $ were one order of
 magnitude larger than the estimates in Eq.~\ndecest. This is true
irrespective of the polarization of
initial and/or final states. In reference \boudjema\
the limits on the $ \alpha $ are determined, the results are
$$ -1.6 < \alpha_2 < 4.8 ; \qquad -0.33 < \alpha_3 < 0.21 \eqn\eq . $$

In reference \dawson\ the effects of the $P$ and $C$ violating (but $ CP $
conserving) term $ \lcal_{ 11 } $ in \ndecops\ were studied for
the cross section and forward-backward asymmetry of $W$ pair
production in polarized
electron positron scattering. Right-handed electron cross section
are found to be much more sensitive to $ \alpha_{ 11 } $ than that
for left-handed electrons. Still the required degree of
polarization must be exceedingly (probably unrealistically)
accurate: $ > $99\%. Even in this case the sensitivity limits
are $ | \alpha_{ 11 } | < 0.2 $.

In reference \frere\ the phenomenological effects of a
modified low energy theory are studied. The light particles are assumed to
be the usual \sm\ excitations together with a neutral heavy vector
boson $Z'$, an $ \su2_L $ scalar singlet $ \chi $, and three right
handed neutrinos. These particles are arranged in an $ \su2 \times
\ui \times \ui $ gauge theory for which the effective operators
are studied. As mentioned in section 2 the number of effective
operators is much larger than the ones presented in Ref.~\bw\
due to the increased particle content. Significant effects on,
$W$ pair production were found  only when the $Z'$ was
relatively light ($ \lesim 300 \gev $), when the couplings
of the effective operators were enhanced by three orders of magnitude
above their natural size. This is true even
when the scale for physics beyond
the $Z'$ is also quite light, of the order of $ 350 \gev $.

\section{{\bfit{LHC.}}}

\item{\bullet}{\it Four-fermi operators}.
A thorough analysis for $pp$ colliders can
be found in Ref.~\ehlq; the best bounds found were for dilepton production
the resulting sensitivity limit being $$ { | \alpha_{ 4 \psi ; vec } |
\over \lat^2 } < 0.07 . \eqn\eq $$ For the \dec, using \wicest,
this gives $ \Lambda \gesim 3.8 \tev $.

\item{\bullet}{\it Triple vector boson couplings}.
In references \boudjema,\falk\ the limits for \tbv\ stemming form $WZ$ and
$W \gamma $ production are studied. With a luminosity of $ 10 / $fb
it is found that
$$ -0.96 < \alpha_2 < 0.93; \qquad -0.07 < \alpha_3 < 0.04 ,  \eqn\eq  $$
which are slightly better than those obtained in Refs.
\baggerdawson, \falk.
These limits are about one order of magnitude larger than the
natural values for these coefficients.

In reference \diakonos\ $ W \gamma +$jet production was considered
obtaining the limits $ -0.04 < \kappa_\gamma - 1 < 0.02 $
and $ | \lambda_\gamma | < 0.04 $. These are about one order
 of magnitude above the estimates \ndecest. For the \dec\ they correspond to
$$ - 1.6 < { \alpha_{ W B } \over \lat^2 } < 0.8 ; \qquad { | \alpha_W |
\over \lat^2 } < 2.5 . \eqn\eq $$

In reference \gounarisi\ the limits expected at LHC
for the coefficients of
the terms $ \lcal_W $ and $ \lcal_{ \phi W } $ are studied
within the \dec. The rationale behind this choice is the assumption
that the underlying physics preserves the custodial symmetry.
The reaction studied is the production of transverse vector bosons
(which in contrasts to the usual \ndec\
scenario~\refmark{\baggerdawson}
is enhanced with respect to the longitudinal vector boson production).
They obtain $$ { | \alpha_W | \over \lat^2 } < 3.6 ; \qquad
{ | \alpha_{ \phi W } | \over \lat^2 } < 7.8 . \eqn\eq $$ The
natural magnitude of these couplings are given in \wicest.

\section{{\bfit{LEP$\times$LHC.}}}

The effects of \tbv\ in this accelerator have been considered in
Ref.~\helbig\
for $ \nu \gamma $ production, and in Ref.~\kim\ for lepton--vector-boson
production. The assumed luminosities vary from $ 500 /$pb to
$ 5000 /$pb and both LEP1-LHC and LEP2-LHC are considered. The
sensitivity limits are all of the same order of magnitude, namely,
$ | \kappa - 1 | \lesim 0.3 ; \ | \lambda | \lesim 0.2 $ which translates
into $$ { | \alpha_{ W B } | \over \lat^2 } ,
{ | \alpha_W | \over \lat^2 } < 12 \eqn\eq $$
assuming the same couplings for  the $Z$ and $ \gamma $ cases.
The bounds are too weak to be of interest in the \ndec.

\section{{\bfit{NLC.}}}

Various versions of this future collider have been considered in the
literature. Two popular choices correspond to energies
of $ 0.5 $ and $ 1 $ \tev\ with luminosities of $ 10 /$fb
and $ 44 /$fb respectively;
these will be referred to as the $0.5 \tev $ and $ 1 \tev $
options.
The laser option~\refmark{\laser} has also been investigated
for the $ e \gamma $ and $ \gamma \gamma $ initial states.

First I consider the effects of the four fermion interactions.
These can be gleaned, for example, from studies of sensitivity limits
of new gauge bosons at the NLC; from Ref.~\djouadi\ I obtain
$$ { | \alpha_{ 4 \psi; vec } | \over \lat^2 } < 0.04 \eqn\eq $$
corresponding to $ \lat > 5 $ when $ | \alpha_{ 4 \psi ; vec } | = 1 $.

For the $ 0.5 \tev $ option, deviations from the \sm\ results for the
reactions $ e^+ e^- \rightarrow W^+ W^- $ and
$ e^+ e^- \rightarrow W^+ W^- V $ ($ V = Z , \gamma $)
generate the limit~\refmark{\boudjema,\likhoded}
$ | \kappa - 1 | \lesim 5 \times 10^{-3 } $,
which is improved by a factor of five
in the $1 \tev $ option.~\refmark{\likhoded} These expectations
are about a factor of four better than those derived in Ref.~
\yehudai;
this last reference also estimates a sensitivity of
$ |\lambda | \le 0.03 $, which corresponds to $ | \alpha_W | < 2 \lat^2 $.
For twice the luminosity the authors of Ref.~\grossekneterschildknecht\
estimate $ | \kappa - 1 | \lesim 0.05 $.
These discrepancies arise from the various observables considered as
well as form the assumptions regarding the observability of the deviations
from the \sm. In this review I will follow the analysis of
Ref.~\barklowi\ which predicts a sensitivity limit of  $
| \lambda | , | \kappa - 1| < 0.01 $, which translates into
$$ { | \alpha_{ W B } | \over \lat^2 } < 0.4 ; \qquad
{ | \alpha_W | \over \lat^2 } < 0.6 , \eqn\eq $$ for the $ 0.5 $ \tev\
option.

The \ndec\ coefficients have also been bounded~\refmark{\boudjema}
with the results, for the $ 0.5 $\tev\ option,
$$ \matrix{ - 0.05 < \alpha_2 < 0.4 , & -0.04 < \alpha_3 < 0.02 , \cr
| \alpha_4| < 0.7 , &  -0.6 < \alpha_5 < 0.5 .
\cr } \eqn\eq $$ For the $ 1 \tev $ option the limits become
$$ - 0.03 < \alpha_2 < 0.1; \qquad | \alpha_3 | < 0.01 . \eqn\eq $$
These are already in the interesting range of a few percent. For higher
energy colliders the sensitivity is improved.~\refmark{\barklow}.
Note however that for a CM energy of $1$\tev, the corrections
from higher index operators will be $ \sim (1 \tev )/( 4 \pi v )
\sim $33\%

A bound on $ \alpha_{ 11 } $ using $W$ pair production for polarized beams
is derived in Ref.~\dawson; the authors consider the
total cross section and the forward-backward asymmetry. The sensitivity
limit in, for example, the $1 \tev$ option are $ | \alpha_{ 11 } |
\lesim 0.01 $ assuming near perfect polarization ($ > $99\%).

In reference \yehudai\ the process $ e \gamma \rightarrow W \nu $ is used
(in the $ 0.5 $\tev\ option)
to obtain the sensitivity bounds $
- 0.07 < \lambda < 0.05 , \
-0.13 < \kappa - 1 < 0.07 $ which correspond to
$$ -4.3  < { \alpha_W  \over \lat^2 } < 3.1 ; \qquad
- 5.1 < { \alpha_{ W B } \over \lat^2 } < 2.7 . \eqn\eq $$
A related investigation~\refmark{\dawson}
considered the effects on the reaction
 $ e \gamma \rightarrow Z W \nu $ generated by the term $ \lcal_{ 11 }
$ in \ndecops. Significant deviations from the \sm\
are found for $ \alpha_{ 11 } = 0.2 $ though no sensitivity limit is
presented.

Finally  the reaction $ \gamma \gamma \rightarrow W^+W^- $ is studied
(in the $ 0.5 $\tev\ option)
deriving the limits $ | \lambda | \le 0.03 $;
$ -0.02 < \kappa - 1 < 0.03 $. These are again
of interest only in the \ndec\ and correspond to
$$ { | \alpha_W | \over \lat^2 } < 1.8 ; \qquad
- 0.8 < { \alpha_{ W B } \over \lat^2 } < 1.2 . \eqn\eq $$
This reaction was also studied in Ref.
\gonzalez\ but the parametrization
used corresponds to a set of dimension eight operators. A consistent
interpretation of their results would require a complete list of
dimension 8 operators relevant for this process; to derive
such a list is a (daunting) task which lies beyond the scope of this
review.

\section{{\bfit{Other experiments.}}}

Various authors have studied the decays of Higgs particles and
their sensitivity to \tbv. In reference \pereztoscano\ the decay
$ H \rightarrow \gamma \gamma $ was studied (see also Ref.~\derujula).
With the constraint~\refmark{\derujula} $ | \alpha_{ W B } | / \lat^2 < 0.7 $
the corresponding width can be up to ten times the \sm\ value so that
this will decay will be a sensitive probe into heavy physics. In other
publications (see for example Ref.~\konig) the same process is considered but
the treatment of divergences and of gauge invariance is incorrect.

In Ref.~\perez\ the effective lagrangian approach is applied to the case
where the low energy fields are the \sm\ plus one additional
doublet. This reference studied the decay $ a \rightarrow \gamma \gamma $,
where $a$ is the CP odd scalar in the model (see Ref.~\gunionetal\ for a
clear review). This process occurs via quark loops, but
these graphs are suppressed
in the limit of large $ \tan \beta $; effective operators can
in this case dominate the process, making it a good candidate
reaction where to look for new physics when (and if) the scalars
used in this model are discovered.

\section{{\bfit{Tables.}}}

In this section I will summarize the results in a series of tables.
The first two present the limits on $ \Lambda $ for the \dec\
obtained using
various operators first for present experiments, then for near
future ones.
The last two tables give the limits on the coefficients $ \alpha $
derived from present and future experiments when $ \Lambda = 1 \tev $.

\def\tinila{$\Lambda $}

\bigskip

{\eightrm
\vbox to  4.5 in{\hsize=4.5 in
\centerline{
\vbox to .5 in{\hsize=3 in \iitem{Table 1.}
Limits on \tinila\ in the \dec\ derived from
existing accelerators. The operators in the left
column generate the corresponding limits.}}
\vbox{\tabskip=0pt \offinterlineskip
\def\tablerule{\noalign{\hrule}}
\halign to 4.5 in{\strut#& \vrule#\tabskip=1em plus 2em&
\hfil#& \vrule#& \hfil#\hfil& \vrule #&
\hfil#& \vrule#& \hfil#\hfil& \vrule #&
\hfil#& \vrule#& \hfil#\hfil& \vrule #\tabskip=0pt\cr\tablerule
\omit&height.1in&\multispan{11}&\cr
&&\multispan{11}\hfil Limits on $ \Lambda $ (in \tev) \hfil&\cr
&&\multispan{11}\hfil (natural size couplings) \hfil&\cr
\omit&height .1in&\multispan{11}&\cr\tablerule
\omit&height.02in&\multispan{11}&\cr\tablerule
&&\omit\hidewidth Operator\hidewidth&&
\multispan{9}\hfil Experiment (present) \hfil&\cr
\tablerule
&&\omit\hidewidth {}\hidewidth &&
\omit\hidewidth AGS821 \hidewidth&&
\omit\hidewidth CLEO \hidewidth&&
\omit\hidewidth HERA \hidewidth&&
\omit\hidewidth LEP1 \hidewidth&&
\omit\hidewidth Tevatron\rlap{$^\dagger$} \hidewidth&\cr\tablerule
&& $\ocal_{ \mu B }$\rlap*
&& 0.5 && --- && --- && --- && --- &\cr\tablerule
&& $\ocal_{ \mu W }$\rlap*
&& 0.5 && --- && --- && --- && --- &\cr\tablerule
&& $\ocal_{ \tau B }$\rlap*
&& --- && --- && --- && 0.01 && --- &\cr\tablerule
&& $\ocal_{ W }$
&& 0.04 && 0.01 && 0.01 && 0.01 && 0.02 &\cr\tablerule
&& $\ocal_{ \tilde W }$
&& --- && 0.01 && --- && --- && --- &\cr\tablerule
&& $\ocal_{ W B }$
&& 0.01 && 0.01 && 0.01 && 0.2 && 0.01 &\cr\tablerule
&& $\ocal_{ \tilde W B }$
&& --- && 0.02 && --- && --- && --- &\cr\tablerule
&& $\ocal_{ 4 \psi; vec }$\rlap{$^\ddagger$}
&& --- && --- && 1.6 && 0.7 && 1.7 &\cr\tablerule
&& $\ocal_{ 4 \psi; sc }$\rlap{$^\ddagger$}
&& --- && --- && 0.4 && --- && --- &\cr\tablerule
&& $\ocal_{ \phi f }^{(3)}$
&& --- && --- && --- && 1.0 && --- &\cr\tablerule
&& $\ocal_{ \phi }^{(1)}$
&& --- && --- && --- && 1.1 && --- &\cr\tablerule
&& $\ocal_{ \phi }^{(3)}$
&& --- && --- && --- && 2.8 && --- &\cr\tablerule
\noalign{\smallskip}
&\multispan{13}* (Yukawa coupling included in coefficient)\hfil\cr
&\multispan{13}$\ddagger$ ($ | \alpha_{4 \psi ; vec, sc } | = 1 $)\hfil\cr
&\multispan{13}$\dagger$ (100/pb)\hfil\cr}}
}}

\bigskip
\bigskip

In table 1 I present the limits on $ \Lambda $ derived from current
experimental data when the coefficients have their natural magnitudes.
As mentioned previously, these estimates have several caveats that
allow situations where the sensitivity is reduced
(small coupling constants or suppression
due to unknown symmetries), or enhanced (low lying resonances).

\bigskip

{\eightrm
\vbox to  3.5 in{\hsize=4.5 in
\centerline{\vbox to .05 in{\hsize=3 in \iitem{{\eightrm
Table 2.}} Limits on \tinila\ in the \dec\ expected from
future accelerators. The operators in the left
column generate the corresponding limits.}}
\vbox{\tabskip=0pt \offinterlineskip
\def\tablerule{\noalign{\hrule}}
\halign to 4.5 in{\strut#& \vrule#\tabskip=1em plus 2em&
\hfil#& \vrule#& \hfil#\hfil& \vrule #&
\hfil#& \vrule#& \hfil#\hfil& \vrule #&\hfil#& \vrule#
\tabskip=0pt\cr\tablerule
\omit&height.1in&\multispan{9}&\cr
&&\multispan{9}\hfil Limits on $ \Lambda $ (in \tev) \hfil&\cr
&&\multispan{9}\hfil (natural size couplings) \hfil&\cr
\omit&height .1in&\multispan{9}&\cr\tablerule
\omit&height.02in&\multispan{9}&\cr\tablerule
&&\omit\hidewidth Operator\hidewidth&&
\multispan{7}\hfil Experiment (future) \hfil&\cr
\tablerule
&&\omit\hidewidth {}\hidewidth &&
\omit\hidewidth LEP2 \hidewidth&&
\omit\hidewidth LHC \hidewidth&&
\omit\hidewidth LEP$\times$LHC \hidewidth&&
\omit\hidewidth NLC\rlap* \hidewidth&\cr\tablerule
&& $\ocal_{ W }$
 && 0.01 && 0.05 && 0.02 && 0.10 &\cr\tablerule
&& $\ocal_{ \tilde W }$
 && 0.01 && --- && --- && --- &\cr\tablerule
&& $\ocal_{ W B }$
 && 0.02 && 0.06 && 0.02 && 0.13 &\cr\tablerule
&& $\ocal_{ \tilde W B }$
 && 0.02 && --- && --- && --- &\cr\tablerule
&& $\ocal_{ 4 \psi; vec }$\rlap{$^\dagger$}
 && 2.8 && 3.8 && --- && 5 &\cr\tablerule
&& $\ocal_{ \phi f }^{(3)}$
 && 4.1 && --- && --- && --- &\cr\tablerule
&& $\ocal_{ \phi W }$
 && --- && 0.03 && --- && --- &\cr\tablerule
\noalign{\smallskip}
&\multispan{11}* (0.5 \tev\ option)\hfil\cr
&\multispan{11}$\ddagger$ ($ | \alpha_{ 4 \psi ; vec } | = 1 $)\hfil\cr}}
}}

\bigskip\bigskip

In many cases the bound on the heavy physics scale lies
below $ \sqrt{ s } $ in collider experiments.
Usually, this means that any new physics within grasp of these
colliders would have been observed directly.
That this is not the case implies that the minimal
value of $ \Lambda $ to be deduced from a given
accelerator is of order $ \sqrt{s} $, many of the above limits
are significantly weaker.
Still quite acceptable bounds can be obtained
from AGS821 and from LEP1; by choosing observables affected by
tree level generated operators, scales of up to a few \tev\ have been
probed.

In table 2 I present the corresponding results for future accelerators.

Again in all but one case the scales probed using effective operators
lie below  $ \sqrt{s } $ and again this implies that if there
is any new physics within reach of these accelerators it will
be produced directly and will not be inferred indirectly
via the low energy effective lagrangian it generates.

The one exception are the expected sensitivity for the four fermion
interactions. In the \ndec\ LHC will probe scales of a few \tev\
in this manner. In the \ndec\ this bound is extended to $ 49 $ \tev;
one must remember that the expected scale in this case is
$ \Lambda  \sim 4 \pi v \simeq 3 \tev $. It must be remembered
that even in this case the determination of new physics effects
is far from straightforward.~\refmark{\bagger}

\bigskip

{\eightrm
\vbox to  6.7 in{\hsize=4.5 in
\centerline{
\vbox to .5 in{\hsize=3 in \iitem{Table 3.}
Limits on the effective lagrangian coefficients from current experiments
when  \tinila =1 \tev.}}
\vbox{\tabskip=0pt \offinterlineskip
\def\tablerule{\noalign{\hrule}}
\halign to 4.5 in{\strut#& \vrule#\tabskip=1em plus 2em&
\hfil#& \vrule#& \hfil#\hfil& \vrule #&
\hfil#& \vrule#& \hfil#\hfil& \vrule #&
\hfil#& \vrule#& \hfil#\hfil& \vrule #\tabskip=0pt\cr\tablerule
\omit&height.2in&\multispan{11}&\cr
&&\multispan{11}\hfil Limits on the coefficients
\hfil&\cr
&&\multispan{11}\hfil (\tinila=1 \tev) \hfil&\cr
\omit&height .2in&\multispan{11}&\cr\tablerule
\omit&height.02in&\multispan{11}&\cr\tablerule
&height.1in&&&\multispan{9}&\cr
&&\omit\hidewidth Coupling\hidewidth&&
\multispan{9}\hfil Experiment (present) \hfil&\cr
&height.1in&&&\multispan{9}&\cr\tablerule
&&\omit\hidewidth {}\hidewidth &&
\omit\hidewidth AGS821 \hidewidth&&
\omit\hidewidth CLEO \hidewidth&&
\omit\hidewidth HERA \hidewidth&&
\omit\hidewidth LEP1 \hidewidth&&
\omit\hidewidth Tevatron\rlap{$\dagger$} \hidewidth&\cr\tablerule
&& $|\alpha_{ 4 \psi; vec }|$
&& --- && --- && 0.4 && 1.9 && 3 &\cr\tablerule
&& $|\alpha_{ 4 \psi; sc }|$
&& --- && --- && 7 && --- && --- &\cr\tablerule
&& $|\alpha_{ \phi f }^{(3)}|$
&& --- && --- && --- && 1.0 && --- &\cr\tablerule
&& $|\alpha_{ \phi }^{(1)}|$
&& --- && --- && --- && 0.9 && --- &\cr\tablerule
&& $|\alpha_{ \phi }^{(3)}|$
&& --- && --- && --- && 0.13 && --- &\cr\tablerule
&& $|\alpha_{ \mu B }|$\rlap*
&& 0.03 && --- && --- && --- && --- &\cr\tablerule
&& $|\alpha_{ \mu W }|$\rlap*
&& 0.03 && --- && --- && --- && --- &\cr\tablerule
&& $|\alpha_{ \tau B }|$\rlap*
&& --- && --- && --- && 1.1 && --- &\cr\tablerule
&& $|\alpha_{ W }|$
&&  4 && 136 && 68 && 37 && 25 &\cr\tablerule
&& $|\alpha_{ \tilde W }|$
&& --- && 57.3 && --- && --- && --- &\cr\tablerule
&& $|\alpha_{ W B }|$
&& 0.2 && 29 && 55 && 0.13 && 59 &\cr\tablerule
&& $|\alpha_{ \tilde W B }|$
&& --- && 12.5 && --- && --- && --- &\cr\tablerule
&& $|\beta_1|$
&& --- && --- && --- && 0.004 && --- &\cr\tablerule
&& $|\alpha_1|$
&& --- && --- && --- && 0.02 && --- &\cr\tablerule
&& $|\alpha_2|$
&& --- && --- && --- && --- && --- &\cr\tablerule
&& $|\alpha_3|$
&& --- && --- && --- && --- && --- &\cr\tablerule
&& $|\alpha_4|$
&& --- && --- && --- && --- && --- &\cr\tablerule
&& $|\alpha_5|$
&& --- && --- && --- && --- && --- &\cr\tablerule
&& $|\alpha_8|$
&& --- && --- && --- && 0.02 && --- &\cr\tablerule
&& $|\alpha_{11}|$
&& --- && 1.8 && --- && 0.9 && --- &\cr\tablerule
\noalign{\smallskip}
&\multispan{13}* (Yukawa coupling included in coefficient)\hfil\cr
&\multispan{13}$\dagger$ (100/pb)\hfil\cr}}
}}

\bigskip
\bigskip

Now I turn to the expected and measured sensitivity to the
effective lagrangian in future and present experiments.
In this case it is of interest to display also the sensitivity for
the coefficients in the \ndec\ as well as in the \dec.  I have
chosen $ \Lambda = 1 \tev $ to present these results. The
translation to other scale is straightforward by using the
formulas presented in the previous sections.

For the \dec\ I again emphasize that, since the scale chosen
lies below $ 4 \pi v $ the contribution from tree level generated
dimension 8 operators could be significant.

\bigskip

{\eightrm
\vbox to 4.5 in{\hsize=4.5 in
\centerline{
\vbox to 0.5 in{\hsize=3 in \iitem{Table 4.}
Limits on the effective lagrangian coefficients
expected from future accelerators when
\tinila =1 \tev.}}
\vbox{\tabskip=0pt \offinterlineskip
\def\tablerule{\noalign{\hrule}}
\halign to 4.5 in{\strut#& \vrule#\tabskip=1em plus 2em&
\hfil#& \vrule#& \hfil#\hfil& \vrule #&
\hfil#& \vrule#& \hfil#\hfil& \vrule #&\hfil#& \vrule#
\tabskip=0pt\cr\tablerule
\omit&height.2in&\multispan{9}&\cr
&&\multispan{9}\hfil Limits on the coefficients
\hfil&\cr
&&\multispan{9}\hfil (\tinila=1 \tev)  \hfil&\cr
\omit&height .2in&\multispan{9}&\cr\tablerule
\omit&height.02in&\multispan{9}&\cr\tablerule
&height.1in&&&\multispan{7}&\cr
&&\omit\hidewidth Coupling\hidewidth&&
\multispan{7}\hfil Experiment (future) \hfil&\cr
&height.1in&&&\multispan{7}&\cr\tablerule
&&\omit\hidewidth {}\hidewidth &&
\omit\hidewidth LEP2 \hidewidth&&
\omit\hidewidth LHC \hidewidth&&
\omit\hidewidth LEP$\times$LHC \hidewidth&&
\omit\hidewidth NLC\rlap* \hidewidth&\cr\tablerule
&& $\alpha_{ W }$
 && 31 && 2.5 && 12 && 0.6 &\cr\tablerule
&& $\alpha_{ \tilde W }$
 && 31 && --- && --- && --- &\cr\tablerule
&& $\alpha_{ W B }$
 && 20 && 1.6 && 12 && 0.4 &\cr\tablerule
&& $\alpha_{ \tilde W B }$
 && 20 && --- && --- && --- &\cr\tablerule
&& $\alpha_{ 4 \psi; vec }$
&& 0.13 && 0.07 && --- && 0.04 &\cr\tablerule
&& $\alpha_{ \phi f }^{(3)}$
 && 0.06 && --- && --- && --- &\cr\tablerule
&& $\alpha_{ \phi W }$
 && --- && 7.8 && --- && --- &\cr\tablerule
&& $\alpha_2$
 && 4.8 && 0.96 && --- && 0.4 &\cr\tablerule
&& $\alpha_3$
 && 0.3 && 0.07 && --- && 0.04 &\cr\tablerule
&& $\alpha_4$
 && --- && --- && --- && 0.7 &\cr\tablerule
&& $\alpha_5$
 && --- && --- && --- && 0.6 &\cr\tablerule
\noalign{\smallskip}
&\multispan{11}* (0.5 \tev\ option)\hfil\cr}}
}}

\bigskip
\bigskip

The measured (or shortly expected) limits on the effective lagrangian
coefficients are presented in table 3 above.
The expected magnitude for these coefficients is $ \sim 1/ 16 \pi^2 \simeq
0.006 $ except for the first five which are naturally of order one
(see \wicest\ and \ndecest). The coefficient $ \beta_1 $ is proportional
to the oblique $T$ parameter and is suppressed, as was
discussed above, for some unknown reason (though it may be understood
on the basis that the heavy physics respects the custodial
symmetry). The coefficients $ \alpha_{ 2 , 8 } $ are also
very well measured being proportional to the oblique $S$ and $U$
parameters. The only other constants which are reasonably well measured
are those expected to be $ \sim 1 $, not coincidentally these also
provide the only significant limits on $ \Lambda $.

As can be seen from the above table, in order to derive
a significant bound on $ \Lambda $ (say, $ \Lambda = 1 \tev $)
from, for example, the Tevatron data by using the anomalous $W$
couplings, one
would have to assume that these couplings are more than four
orders of magnitude above their natural magnitudes.

In table 4 I present the expected sensitivity of future
colliders, should new physics be at $ 1 \tev $.
As can be seen the expected sensitivity to the
various coefficients is very weak except for
some \ndec\ couplings, especially $ \alpha_3 $
(the same will be true for $ \alpha_\phi \up 3 $). I have not included
the couplings corresponding to the oblique parameters $ \alpha_{ 2 , 8 } $
and $ \beta _1 $ in this study, but the experimental precision will
certainly improve for these measurements, to the point that we must
either see some deviations form the \sm\ or else postulate some
mechanism that will hide the heavy physics effects from the oblique
parameters.

\chapter{Conclusions.}

In the previous sections I have presented a scrutiny of various aspects
of effective theories and their applications to the weak interaction
phenomenology.

Effective theories are a useful instrument for parametrizing
new physics effects in a consistent and process and model independent
manner. The formalism generates corrections to the \sm\
contributions to any observable quantity in terms
of a series with unknown coefficients which
are not expected to be fundamental,
but combination of (as yet unknown) new constants of the
lagrangian describing the underlying physics. Due to the
hierarchy inherent in the effective theories considered, a finite number
of (very high precision) experimental data points would determine the
whole of the new physics effects to a certain accuracy; the errors are
then also estimated using the formalism.

The consistent application of the effective lagrangian
parametrization yields very good limits on the scale of new physics
(up to several \tev\ in some cases). Many experimental constraints
are, however, very weak implying only that
$ \Lambda $ greater than a few \gev. In obtaining these results
it is paramount that the coefficients of the effective lagrangian
should take reasonable values, for example, $ \lambda_Z $ in \tbv\
{\it cannot} be $ O ( 1 ) $ if the approach is to be consistent.

If we adopt the optimistic
assumption that, should there be new physics at a scale $ \Lambda $
all cross sections will show drastic changes at this scale, then we can
use the known bound on $ W_R $ masses from CDF [\partdatbook]
to state that $ \Lambda \gesim 0.5 \tev $ for the \dec.
This, though better than many
limits obtained in the previous section, can of course
be avoided depending on the
type of new physics present. But it still a fact that most limits on
$ \Lambda $ to be achieved by present colliders are in this ball park.
To improve these results colliders of energy significantly
larger than the present
ones are required or, alternatively, super-high precision measurements
(at the one per mil or 0.1 per mil level) are needed. Of these possibilities
only the first will determine the kind of new physics present
unambiguously. For example, it would be very difficult, if not impossible,
to state that a certain deviation in the anomalous magnetic moment of the
muon is due to the presence of an anomalous vector boson coupling and
not to a slight miscalculation of the QCD effects [\martypion].

Any calculation of the effects of new physics via an effective
lagrangian (as opposed to a specific model calculation) must al least
be consistent. The implications of this obvious fact, however,
are often ignored or forgotten. Two such examples were detailed above:

\iitem{{(i)}} One cannot arbitrarily ignore some of the operators. In
particular one cannot include in a calculation loops containing effective
operators and forget others which contribute at tree level, unless
specific assumptions regarding the underlying physics are made which
imply that the tree level contributions are negligible or absent.

\iitem{{(ii)}} The results of any calculation are reliable as long as
the contributions of the operators which were ignored
are significantly smaller than the contributions  from the operators which
were retained.

If these facts are ignored one can obtain
impressive effects which would,
apparently, indicate enormous deviations from the \sm. These results
would be wrong: they were obtained under circumstances
where the effective lagrangian parametrization is invalid.
If  the calculations are performed with the consistency of the
parametrization in mind, then the effects are generally small,
though not always unmeasurable.
For example, processes affected by
four-fermi operators produced by a strongly interacting theory would produce
measurable deviations from the \sm\ at the LHC provided $ \Lambda $ lies
below $ \sim 47 \tev $.

As discussed in section 4.4 one cannot strongly violate the
estimates for the coefficient of $ \leff $ without losing
all predictability in the model. Even if the use  of
the theory in loop calculations is forbidden, the rationale
behind the inter-relations
among the lagrangian coefficients is lost. For example,
instead of understanding the universal lepton couplings to the
$W$ as a consequence of gauge invariance and the representations
carried by the fermions, it becomes an accidental fact.

When the formalism is applied consistently it remains a powerful and
useful tool with which to probe, in a model independent manner, physics
about which we as of now know very little. Unfortunately it is clear from
the results of the last section, that most experiments are sensitive
to only a small number of effective interactions. This necessarily
limits our ability to probe new physics, for it is possible to imagine
situations in which these operators are suppressed. The extraction
of more information remains a difficult and arduous task which
necessitates much more data (to improve statistics) or, optimally,
higher energies.
\ack

It is a pleasure to acknowledge the help of M. Einhorn from whom I learned
most of the ideas presented in the paper.
I also received much help from C. Arzt and M. Golden.
I benefited from many discussion with M.A. Perez, U. Baur, D.
London and E. Ma. I would also like to thank the Aspen Center for Physics
for its hospitality where part of this work was completed.
This work was supported in part by an SSC fellowship,
by a UC Mexus fellowship and through funds provided by the Department of
Energy.

\Appendix{A}

In this appendix I will present, for the \dec, the contributions
to the three vector boson interactions derived from the  operators
of dimension six in the effective lagrangian. This exercise will
be useful since many authors have studied the limits put on these
interactions by the current experimental data, as well as the expected
sensitivity in future colliders. I will use the notation of Ref. \bw.

There are six operators which generate couplings among three vector bosons,
they are
$$ \eqalign{
\ocal_W &= \epsilon_{ I J K } W_{ \mu \nu }^I
W_{ \nu \rho }^J W_{ \rho \mu }^K \cr
\ocal_{ \tilde W} &= \epsilon_{ I J K } \tilde W_{ \mu \nu }^I
W_{ \nu \rho }^J W_{ \rho \mu }^K \cr
\ocal_{ \phi W } &=  \half \left( \phi^\dagger \phi \right)
\left( W_{ \mu \nu }^I \right)^2 \cr
\ocal_{ \phi B } &= \half \left( \phi^\dagger \phi \right)
\left( W_{ \mu \nu }^I \tilde W_{ \mu \nu }^I \right) \cr
\ocal_{ W B } &= \phi^\dagger \sigma^I \phi W_{ \mu \nu }^I B_{ \mu \nu } \cr
\ocal_{ \tilde W B } &= \phi^\dagger \sigma^I \phi
\tilde W_{ \mu \nu }^I B_{ \mu \nu } \cr } \eqn\eq $$
where $ B_{ \mu \nu } = \partial_\mu B_\nu - \partial_\nu B_\mu $,
$ W_{ \mu \nu }^I = \partial_\mu W_\nu^I - \partial_\nu W_\mu^I
+ g \epsilon_{ I J K } W_\mu^J W_\nu^K  $ in which $ W_\mu^I $
and $ B_\mu $ denote the $ \su2_L $ and $ \ui_Y $ gauge fields
respectively; the corresponding gauge couplings are
$g$ and $ g' $. The scalar doublet is denoted by
$ \phi $ and $g$ is the $ \su2_L $ gauge coupling constant.
Note that all these operators are gauge invariant.

In the unitary gauge $ \phi $ is replaced by its \vev; in this
case the effects of $ \ocal_{ \phi W } $ and $ \ocal_{ \phi \tilde W } $
can be absorbed in a redefinition of the \sm\ lagrangian parameters.
These operators will not be considered further. The effective
lagrangian of interest then becomes $$ \eqalign{
\leff = \lcal \lowti{St. Model} +
& \inv{ \Lambda^2 } \left[
g^3 \alpha_W \ocal_W + g^3 \alpha_{ \tilde W } \ocal_{ \tilde W } + \right.
\cr & \qquad \qquad \left.
g' g \alpha_{ W B } \ocal_{ W B }
+ g' g
\alpha_{ \tilde W B } \ocal_{ \tilde W B } \right] . \cr } \eqn\lefftbv $$

Using the results of section 4.2 I can estimate the
natural magnitude for the above coefficients. the results are
$$ \alpha_{ W , \tilde W , W B , \tilde W B }
\sim \inv{ 16 \pi^2 } . \eqn\eq $$

{}From \lefftbv\ one can easily derive all triple boson vertices generated
by the dimension six operators. All such couplings are of the form
$ W^+ W^- V $ with $ V = Z , \ \gamma $. The $WWZ$ couplings are
$$ \eqalign{
{ \Lambda^2 \over e \; \cot \theta_W } \lcal_{ W W Z } =&
6 i g^2 W^-_{\nu \rho } W^+_{ \rho \mu } \left( \alpha_W
Z_{ \mu \nu }  + \alpha_{ \tilde W } \tilde Z_{ \mu \nu } \right) \cr
& + 4 i m_W^2 W^-_\mu W^+_\nu \left( \alpha_{ W B } Z_{ \mu \nu }
+ \alpha_{ \tilde W B } \tilde Z_{ \mu \nu } \right) , \cr } \eqn\eq $$
while the $WW\gamma $ couplings are
$$ \eqalign{
{ \Lambda^2 \over e } \lcal_{ W W \gamma } =&
6 i g^2 W^-_{\nu \rho } W^+_{ \rho \mu } \left( \alpha_W
Z_{ \mu \nu }  + \alpha_{ \tilde W } \tilde Z_{ \mu \nu } \right) \cr
& - 4 i m_W^2 W^-_\mu W^+_\nu \left( \alpha_{ W B } Z_{ \mu \nu }
+ \alpha_{ \tilde W B } \tilde Z_{ \mu \nu } \right) . \cr } \eqn\eq $$

Recall that there are 14 \  $WWZ$ and $ WW \gamma $
lorentz invariant (though not manifestly gauge invariant) couplings
in \tbv.
These are reduced to four within this formalism. If \tbv\
is made gauge invariant
along the lines of section 2, then the resulting lagrangian mixes
terms of different indices (see section 3 for the definition of
the index of an operator). Keeping only the terms of the lowest index
again results in a reduction in the number of unknown parameters.

Finally note that the operators $ \ocal_W $ and $ \ocal_{ \tilde W } $
also generate four-vector boson couplings, and it is easy to see that all
such vertices involve at least two $W$ fields. It is  also important to
note that there are no other sources of four vector boson couplings in
the operators of dimension six (these are also the only operators
generating couplings among five and six vector bosons).
This implies that all couplings among three vector bosons
or more depend on four parameters only. If CP violations are assumed to
be negligible the number of unknown parameters is halved.

\Appendix{B}

For convenience I include the expressions for the operators of index
two in the \ndec, the results are taken from Ref. \appelquistwu. The
notation used is $$ T = U^\dagger \sigma_3 U ; \qquad V_\mu = \left(
D_\mu U \right) U^\dagger ; \qquad \WW_{ \mu \nu } = \half W_{ \mu \nu }^I
\sigma_I , \eqn\eq $$ where $ D_\mu  U = \partial_\mu U + { i \over
2 } g \sigma_I W_\mu^I U - { i \over 2 } g' B_\mu U \sigma_3 $.

The CP conserving terms are $$ \eqalign{
\lcal_1 &= \half \alpha_1 g g' \; B_{ \mu \nu } \tr( T \; \WW^{ \mu
\nu } ) , \qquad
\lcal_2 = \half i \alpha_2 g' \; B_{ \mu \nu } \tr( T \; [ V^\mu ,
V^\nu ] ) , \cr
\lcal_3 &= i \alpha_3 g \; \tr( \WW_{ \mu \nu } [ V^\mu , V^\nu ] ) ,
\qquad
\lcal_4 = \alpha_4 \left\{ \tr ( V_\mu \; V_\nu ) \right\}^2 , \cr
\lcal_5 &= \alpha_5 \left\{ \tr ( V_\mu \; V^\mu ) \right\}^2 ,
\qquad \qquad
\lcal_6 = \alpha_6 \; \tr(V_\mu \; V_\nu ) \; \tr( T \; V^\mu )
\; \tr( T \; V^\nu ) , \cr
\lcal_7 &= \alpha_7 \; \tr( V_\mu V^\mu ) \; \left\{ \tr( T \;
V^\nu ) \right\}^2 , \qquad
\lcal_8 = \quarter \alpha_8 g^2 \left\{ \tr ( T \; \WW_{ \mu \nu }
) \right\}^2 , \cr
\lcal_9 &= \half i \alpha_9 g \; \tr ( T \; \WW_{ \mu \nu }) \;
\tr( T \; [ V^\mu , V^\nu ] ) , \cr
\lcal_{10} &= \half \alpha_{ 10 } \left\{ \tr( T \; V^\mu )
\; \tr( T \; V^\nu ) \right\}^2 , \cr
\lcal_{ 11 } &= 2 \alpha_{ 11 } g \; \tr( T \; V^\mu ) \; \tr (
V_\nu \; \tilde \WW^{ \mu \nu } ) . \cr } $$

The CP violating operators are $$ \eqalign{
\lcal_{ 12 } &= \alpha_{ 12 } g \;\tr( T \; V_\mu ) \; \tr( V_\nu
\; \WW^{ \mu \nu } ) , \qquad
\lcal_{ 13 } = 2 \alpha_{ 13 } g g' \; \tilde B^{ \mu \nu } \;
\tr( T \; \WW_{ \mu \nu } ) , \cr
\lcal_{ 14 } &= 4 i \alpha_{ 14 } g' \; \tilde B^{ \mu \nu } \;
\tr( T \; V_\mu V_\nu ) , \qquad
\lcal_{ 15 } = 4 i \alpha_{ 15 } g \; \tr( \tilde \WW_{ \mu \nu } \;
V^\mu V^\nu ) , \cr
\lcal_{ 16 } &= \alpha_{ 16 } g^2 \; \tr( T \; \tilde \WW_{ \mu
\nu} ) \; \tr ( T \; \WW^{ \mu \nu } ) , \cr
\lcal_{ 17 } &= i \alpha_{ 17 } g \; \tr( T \; \tilde \WW_{ \mu
\nu} ) \; \tr ( T \; V^\mu V^\nu ) , \cr
\lcal_{ 18 } &= 2 \alpha_{ 18 } g'{}^2 \; \tilde B^{ \mu \nu }
B_{ \mu \nu } , \qquad
\lcal_{ 19 } = \alpha_{ 19 } g^2 \; \tr( T \; \tilde \WW_{ \mu
\nu} ) \; \tr( T \; \WW^{ \mu \nu } ) , \cr } \eqn\eq $$
where $ \tilde B_{ \mu \nu } = \half \epsilon_{ \mu \nu \rho \sigma }
B^{ \rho \sigma } $ and similarly for $ \tilde \WW $.

Finally the operators containing two fermions and index zero, and which
do not correspond to a kinetic or mass terms are $$
\bar \ell_l U \sigma_3 U^\dagger \ell ; \qquad
\bar \ell U^\dagger \left( \not \!\! D U \right) \ell ; \qquad
\tr \left( T \; V_\mu \right) \bar l_R \gamma_\mu l_R ; \eqn\eq $$
plus its counterparts for $q$, $u_R $ and $ d_R $.

\Appendix{C}

In this appendix a simple example will be provided to illustrate the
consequences of having a large number of particles in a loop in order
to offset the loop suppression factors described in section 4.4.

The example consists of a simple two dimensional model of $N$ heavy
fermions interacting with a light scalar field $ \theta $. The lagrangian
is $$ \lcal = \half \left( \partial \theta \right)^2
+ \sum_{ a = 1 }^N  \bar \psi_a i \dcal \psi_a;
\qquad \dcal = \not \! \partial + { i g \over 2 }
\left( \partial_\mu \theta \right) \gamma^\mu
- i m e^{ i g \theta
\gamma_5 } . \eqn\eq $$
This is invariant under $ \theta \rightarrow \theta - \alpha $,
$ \psi \rightarrow \exp ( i \alpha g \gamma_5 /2 ) \psi $.

The fermionic path integral can be evaluated by noting that the scalar
field can be removed from the fermionic terms by a chiral rotation.
Let $$ W ( \theta ) = \int \left[ d \psi d \bar \psi \right]
\exp \left( i \int d^2 x \; \bar \psi i \dcal \psi \right ) $$
then the change $ \theta \rightarrow \theta - \alpha $, for
$ \alpha $ infinitesimal,
accompanied by a change of variables
$ \psi \rightarrow ( 1 + i \alpha g \gamma_5 /2 ) \psi $
leaves the lagrangian invariant.
There is, however, a Jacobian which is evaluated using Fujikawa's
technique~\refmark{\fujikawa}. The quantity to be evaluated is
$ \tr \gamma_5 \exp \left[ - \dcal^2 / M^2 \right] $ in the limit
$ M \rightarrow \infty $; a straightforward
calculation gives $$
W ( \theta - \alpha ) = W ( \theta )-   \int { d^2 x \over 4 \pi }
\left[ g^2 ( \partial^\mu \theta ) ( \partial_\mu \alpha )
+ m^2 ( 2 \alpha g ) \sin ( 2 g \theta ) \right] ,\eqn\eq $$
whence $$ W ( \theta ) = \int { d^2 x \over 4 \pi } \left[
\half g^2 ( \partial \theta ) ^2 - m^2 \; \cos (2 g \theta ) \right]
\eqn\eq $$

The full effective lagrangian for $ \theta $ when $N$ heavy fermions
are integrated out is
$$ \leff = \half \left(
1 + { g^2 N \over 4 \pi } \right)
( \partial \theta )^2 - { N m^2 \over 4 \pi }
\cos ( 2 g \theta ) . \eqn\eq $$
The quantity $ g^2 N / ( 4 \pi ) $ represents the loop factor in this case.

Let $$ \lambda = { 2 g \over
\sqrt{ 1 + g^2 N / ( 4 \pi ) } } , \qquad
c = \sqrt{ 1 + { 4 \pi \over g^2 N } } ; \eqn\somedefs
$$ then a field redefinition
$ \chi = ( \pi + 2 g \theta ) / \lambda $ gives
$$ \leff = \half ( \partial \chi )^2 + { m^2 N \over 4 \pi } \cos (
\lambda \chi ) \eqn \eq $$

The mass of the linear excitations of this Sine-Gordon
effective lagrangian
(for a review see Ref. \rajaraman)
equals $ 2 m /c $ (where $c$ is defined in \somedefs).
The soliton mass equals $ m N c / \pi $. By definition $ c \ge 1 $,
which implies that the soliton masses are always large, $ O ( m ) $;
the same will be true for all other
``topological'' objects in the model, such
as breathers, etc.
The linear excitations will be light only if $ c \gg 1 $ which is equivalent
to $ 4 \pi \gg g^2 N $.

If the number of fermion loops is so large as to offset
the loop factor $ g^2 / ( 4 \pi ) $ there are no light
excitations at all. If $g$ is assumed to decrease with $N$,
$ g = G / \sqrt{ N } $, then  $ c = \sqrt{ 1 + 4 \pi / G^2 } $ and
the soliton masses are linear in $N$.

\Appendix{D}

The proof that the equations of motion can be used to reduce
the number of operators can be given in general~\refmark{\arztiii}
(see also Ref. \georgi). Here I will present a simple
proof for a non-gauge theory, the above references should be consulted
for the general case.

I will denote the fields by $ \chi $ and the classical action by
$ S ( \chi ) $. Suppose now that we have two operators $ \ocal $
and $ \ocal' $ such that $$ \ocal ' = \ocal + \int d^4 x\; \acal
{ \delta S \over \delta \chi } \eqn\eq $$ for some local quantity
$ \acal $ depending on the $ \chi $. The effective action is
$$ S \lowti{ eff } = S + \int d^4 x \; \left( \eta \ocal +
\eta' \ocal' \right) + \cdots ; \eqn\eq $$  the dots indicate higher
dimensional operators. Let $ S ' = S + \int d^4 x ( \eta + \eta' ) \ocal $,
then $$ \eqalign{ S \lowti{eff }
&= S' ( \chi ) + \eta' \int d^4 x\;
\acal { \delta S \over \delta \chi } + \cdots \cr
&= S' ( \chi ) + \eta' \int d^4 x\;
\acal { \delta S' \over \delta \chi } + \cdots \cr
&= S' ( \chi + \eta' \acal ) + \cdots \cr } \eqn\eq $$
Thus, to the order we are working, the effects of $ \ocal' $ are
to replace $ \eta \rightarrow \eta + \eta' $ and $ \chi \rightarrow
\chi + \eta' \acal $.

The natural next step is to change variables $ \chi \rightarrow \phi
= \chi + \eta' \acal $. The Jacobian is one since $ \acal $ is a local
object and so $ \delta \acal ( x ) / \delta \chi ( y ) $
will be proportional to $ \delta ( x - y ) $ or one of its derivatives.
But in dimensional regularization all these quantities vanish
when $ x = y $, and so there is no contact term. In other regularization
prescriptions the Jacobian will cancel some contact terms
generated by $S'$.

To find the effects of the above change of variables I
use an argument presented in Ref. \coleman.
When considering Green's functions the replacement
$ \chi \rightarrow \phi $ generates many extra terms, especially
if $ \acal $ contains terms with several fields. But the
contributions of the terms in $ \acal $ with more than one
field to any Green function
do not have the physical particle propagator poles. Hence,
when these terms are multiplied by the inverse propagators and the
mass-shell condition is imposed, they vanish.

The only remaining terms are those where $ \acal $ is linear in the
fields  $ \chi $. When these objects are close to the mass
shell their effect is to multiply the $ \eta' = 0 $ contributions
by a finite factor. This precise
same factor will also appear in the propagator.
Therefore when these terms
are multiplied by inverse propagators and put on the mass-shell
all the remaining effects of $ \acal $ cancel.

{}From this discussion it follows that the only S-matrix
effect of adding a redundant
operator to $ \leff $ is to shift the couplings of the existing terms.
This shift, as explained in section 5, can have a very important
quantitative effect. If a consistent parametrization of Green functions
is desired, then  the full set of operators (equivalent or not)
must be included in the effective lagrangian.

\refout

\bye